\documentclass[letter,scriptaddress,pra,showkeys]{revtex4}
\usepackage{amsmath}
\usepackage{makeidx}
\usepackage{amsfonts}
\usepackage[ansinew]{inputenc}
\usepackage[usenames,dvipsnames]{pstricks}
\usepackage{subfigure}
\usepackage{esint}
\usepackage{tikz}
\usepackage{graphicx}

\newcommand{\nn}{\nonumber}
\newcommand{\be}{\begin{equation}}
\newcommand{\ee}{\end{equation}}
\newcommand\bea{\begin{eqnarray}}
\newcommand\eea{\end{eqnarray}}
\newcommand{\bc}{\begin{center}}
\newcommand{\ec}{\end{center}}

\makeindex

\begin{document}

\author{Zhongyou Mo}
\affiliation{School of Physics and Technology, Wuhan University, Wuhan, 430072, China}
\affiliation{Center for Theoretical Physics, Wuhan University, 430072, China}

\author{Junji Jia}
\email[Email: ]{junjijia@whu.edu.cn}
\affiliation{School of Physics and Technology, Wuhan University, Wuhan, 430072, China}
\affiliation{Artificial Micro- and Nano-structures, Wuhan University, 430072, China}

\title{Generalized Schl\"omilch's formulas and thermal Casimir effect of a fermionic rectangular box}
\date{today}

\begin{abstract}
Schl\"omilch's formula is generalized and applied to the thermal Casimir effect of a fermionic field confined a three-dimensional rectangular box. The analytic expressions of the Casimir energy and Casimir force are derived for arbitrary temperature and edge sizes. The low and high temperature limits and finite temperature cases are considered for the entire parameter space spanned by edge sizes and/or temperature. In the low temperature limit, it is found that for typical rectangular box, the effective 2-dimensional parameter space spanned by the two edge size ratios can be split into four regions. In one region, all three forces between three pairs of faces are attractive, and in another two regions, the force along the longest edge becomes repulsive and in the last region the force along both the longest and medium sized edge becomes repulsive. Three forces cannot be made simultaneously repulsive. For the waveguide under low temperature, the Casimir force along the longer side of the waveguide cross-section transforms from attractive to repulsive when the aspect ratio of the cross-section exceed a critical value. For the parallel plate scenario under low temperature, our results agrees with previous works. For high tempera limit, it is shown that both the Casimir energy and force approach zero due to the high temperature suppression of the quantum fluctuation responsible for the Casimir energy. For the finite temperature case, we separate the parameter space into four subcases (C1 to C4) and various edge size and temperature effects are analyzed. In general, we found that in all cases the Casimir energy is always negative, while the Casimir force at any finite or low temperature can be either repulsive or attractive depending on the sizes of the edges. For the case (C1) that is similar to parallel plates with relatively high temperature, it is found that the Casimir force is always attractive, regardless the change of the plate separation. At the given temperature, The Casimir energy/force densities approach the infinite parallel plate limit even when the plate edge size is 2 times the plate separation. For the case (C2) that is similar to a waveguide with relatively high temperature, the Casimir force along the longer side of the waveguide cross-section transform from attractive to repulsive when this side exceed a critical value. This critical point forms a boundary in the parameter space when the shorter edge of the waveguide cross-section changes and the boundary values decreases with respect to temperature increase. Case (C3) covers the low temperature parallel plate, typical rectangular box and waveguide geometries. For the waveguide case, the force along the waveguide longitude also transform from attractive to repulsive when the waveguide length exceed certain critical value. These critical values changes with respect to temperature in a nontrivial way. For the typical waveguide case (C4) at low temperature, the Casimir energy density along the longitudinal direction is a constant while force density decrease linearly as the waveguide length increases. Finally, for any fixed temperature, there exist a boundary in the parameter space of edge sizes separating the attractive and repulsive regions.  Besides, the Casimir energy for an electromagnetic field confined in a three-dimensional box is also derived.
\end{abstract}


\maketitle

\section{Introduction}
First proposed in 1948~\cite{Casimir:1948dh}, Casimir effect has been studied extensively using both experimental and theoretical approaches. In the simplest case, Casimir effect is known as an attraction (or repulsion) between two parallel conducting plates due to the fluctuations of vacuum energy. Experimentally it has been observed using different materials, geometries and measurement setups \cite{Sparnaay:1958wg,Arnold,Lamoreaux:1996wh,Mohideen:1998iz}. Theoretically, it is usually studied according to the geometry and boundary conditions, temperature and nature of fields.

Besides the usual parallel plates geometry, other geometries such as cylindrical, spherical boundaries, rectangular cavities and spherical-plate geometries are often studied. In particular, the study of spherical boundaries first by Boyer~\cite{Boyer:1968uf} and later by Milton et al.~\cite{Milton:1978sf} for electromagnetic field showed that the Casimir force could be repulsive too. The theory of Casimir effect for systems with boundaries of real body was established by Lifshitz in Ref. \cite{Lifshitz:1956zz}, where he also considered the effect of temperature.

Temperature is another important factor influencing Casimir effect. The thermal Casimir effect was calculated for electromagnetic and/or scalar field confined in rectangular cavities~in Ref. \cite{Ambjorn,Santos:2000be,Jauregui,Lim,Geyer:2008wb}. Lim and Teo studied the Casimir effect for massless scalar field and electromagnetic field \cite{Lim:2008tc, Lim:2008rj} for piston geometries. Lin and Zhai discussed the finite temperature Casimir effect in general $p$ dimensional rectangular cavity \cite{Lin:2014lva}. Finite temperature Casimir effect for electromagnetic field with a boundary of a spherical shell was computed by Balian and Duplantier~\cite{Balian}, giving the free energy in low and high temperature limits.

The Casimir effect also depends crucially on the nature of the field, i.e., scalar, fermionic, gauge field and mass of the field. In particular, the fermionic field Casimir effect are considered by a series of paper. The Casimir effect for massless Dirac field confined between two parallel plates were studied by Johnson~\cite{Johnson} and Milonni in Ref. \cite{Milonni:1994xx}, where they showed that the Casimir force is attractive as in the case of electromagnetic field. Calculations by Gundersen and Ravndal~\cite{Gundersen:1987wz} showed the Casimir force becomes repulsive at sufficiently high temperatures for massless fermions also confined between parallel plates. In this work, many interesting properties such as temperature inversion symmetry, energy-momentum tensor and fermion condensate were discussed. The Casimir energy for a massless fermionic field confined in a three-dimensional rectangular box at zero temperatures was studied by Seyedzahedi et al.~\cite{Seyedzahedi:2010fya}, showing the Casimir energy is negative as opposed to the case of a three-dimensional sphere considered by Milton~\cite{Milton2} where the Casimir energy is positive. Besides, extra dimension corrections for a three-dimensional box with massless fermionic field were considered by Sukamto and Purwanto~\cite{Sukamto}.

In the present paper, we extend the above works by study the thermal Casimir effect at arbitrary temperature for a massless fermionic field confined in a rectangular box. In doing this, we used a new method, the generalized Schl\"omilch's formulas, for the evaluation of the frequency summation. We also used this method to study the thermal Casimir effect of an electromagnetic field confined a three-dimensional rectangular box and found that the resulting Casimir energy in a cube at zero temperature agrees perfectly with previously reported result at low temperature \cite{Geyer:2008wb}.

This paper is organized as the following. In Sec.~\ref{series} the Schl\"omilch's formula is briefly introduced and generalized to the cases of double series and triple series. In Sec.~\ref{EM field}, the generalized Schl\"omilch's formulas are applied to the thermal Casimir effect of an electromagnetic field confined in a rectangular box. In Sec.~\ref{Fermion field}, the thermal Casimir effect is considered for a massless fermionic field confined in a rectangular box with M.I.T bag model boundary condition. The general formulas of the Casimir energy and force for arbitrary temperature and edges sizes are derived in this section. Then in Sec. \ref{secparaspace} the Casimir effect in the entire parameter space spanned by the temperature and three edge sizes is thoroughly studied, in both analytical and numerical ways. Section \ref{secdis} summarizes the findings and outlines potential extensions of the work and other possible applications of the generalized Schl\"omilch's formula.

\section{Schl\"omilch's formula and its generalization\label{series}}
An useful formula first discovered by Schl\"omilch~\cite{Schlomilch,Schlomilch2} and used in many works \cite{J. Lagrange,Grosswald,Dowker:2002ax} (see \cite{notebook} for older papers) is the following
\be
\alpha\sum_{k=1} \frac{k}{e^{2\alpha k}-1}+\beta\sum_{k=1} \frac{k}{e^{2\beta k}-1}=\frac{\alpha+\beta}{24}-\frac{1}{4}. \label{eq1}
\ee
where $\alpha,\beta>0$, $\alpha\beta=\pi^2$ and the sum here and after runs to infinity until otherwise explicitly specified.
A formula derived from Eq.~\eqref{eq1} which is also useful by itself, is \cite{notebook}
\be
\sum_{k=1} \ln\bigl(1-e^{-\alpha k}\bigl)=\sum_{k=1}\ln\bigl(1-e^{\frac{-4\pi^2 k}{\alpha}}\bigr)-\frac{\ln\alpha}{2}-\frac{\pi^2}{6\alpha}+\frac{\alpha}{24}+\frac{\ln{(2\pi)}}{2}.\label{one-partition}
\ee
The similarity between the Bose-Einstein distribution and terms in Eq.~\eqref{eq1} enables its possible applications in physics, particularly in Casimir effects. It is observable the functions in the sums on the left side of Eq.~\eqref{eq1} look like the average energy $u_A$ of a single resonator in Planck's law for the energy spectrum~\cite{Planck:1901tja}
\be
\text{u}(\nu,T)=\frac{8\pi\nu^2}{c^3}u_A=\frac{8\pi\nu^2}{c^3}\cdot\frac{h\nu}{e^{h\nu/(k_B T)}-1},\label{Planck's law}
\ee
where $T$ is the temperature and $k_B$ is Boltzmann constant.
Because of this, equations~\eqref{eq1} and~\eqref{one-partition} are useful to calculate the internal energy $U$ and free energy $F$ of a one dimensional linear harmonic oscillators system with discrete frequencies~\cite{Feynman}
\bea
U&=&\sum_n\biggl[ \frac{\hbar\omega_n}{2} +\frac{\hbar\omega_n}{e^{\hbar\omega_n/(k_B T)}-1} \biggr],\label{internal-energy}\\
F&=&\sum_n\biggl[ \frac{\hbar\omega_n}{2} +k_B T\ln\Bigl(1-e^{\hbar\omega_n/(k_B T)}\Bigr) \biggr].\label{free-energy}
\eea
For a three-dimensional linear harmonic oscillator system, the series in Eqs.~\eqref{internal-energy} and ~\eqref{free-energy} will contain more than one summation. Therefore, the generalizations of Eqs.~\eqref{eq1} and~\eqref{one-partition} to the cases of double series and triple series are necessary for the purpose of application in Casimir effect.

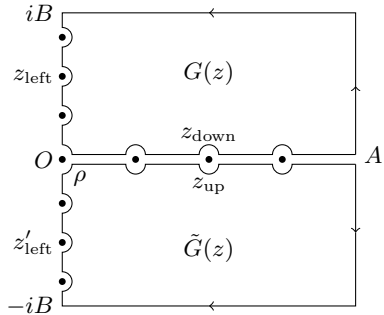
\begin{figure}[htp]
\begin{tikzpicture}[scale=0.65]%scale
\draw(0,0.3)arc(90:0:0.2)--(1.3,0.1)arc(180:0:0.2)--(2.8,0.1)arc(180:0:0.2)--(4.3,0.1)arc(180:0:0.2)--(6,0.1)node[right]{$A$};
\draw[->](6,0.1)--(6,1.5);
\draw[->](6,1.5)--(6,3)--(3,3);
\draw(3,3)--(0,3)node[left]{$iB$}--(0,2.7)arc(90:-90:0.2)--(0,1.9)arc(90:-90:0.2)--(0,1.1)arc(90:-90:0.2)--(0,0.3);
%point
\fill(0,0)circle[radius=2pt]node[left]{$O$};
\fill(0.05,-0.05)circle[radius=0.01pt]node[below right]{$\rho$};
\fill(1.5,0)circle[radius=2pt];
\fill(3,0)circle[radius=2pt];
\fill(4.5,0)circle[radius=2pt];
\fill(0,0.9)circle[radius=2pt];
\fill(0,1.7)circle[radius=2pt]node[left]{$z_{\rm left}$};
\fill(0,2.5)circle[radius=2pt];
\draw(0,-0.3)arc(-90:0:0.2)--(1.3,-0.1)arc(-180:0:0.2)--(2.8,-0.1)arc(-180:0:0.2)--(4.3,-0.1)arc(-180:0:0.2)--(6,-0.1);
\draw[->](6,-0.1)--(6,-1.5);
\draw[->](6,-1.5)--(6,-3)--(3,-3);
\draw(3,-3)--(0,-3)node[left]{$-iB$}--(0,-2.7)arc(-90:90:0.2)--(0,-1.9)arc(-90:90:0.2)--(0,-1.1)arc(-90:90:0.2)--(0,-0.3);
\fill(0,-0.9)circle[radius=2pt];
\fill(0,-1.7)circle[radius=2pt]node[left]{$z'_{\rm left}$};
\fill(0,-2.5)circle[radius=2pt];
\fill(3,1.35)circle[radius=0.01pt]node[above]{$G(z)$};
\fill(3,-1.35)circle[radius=0.01pt]node[below]{$\tilde{G}(z)$};
\fill(3,0.2)circle[radius=0.01pt]node[above]{$z_{\rm down}$};
\fill(3,-0.2)circle[radius=0.01pt]node[below]{$z_{\rm up}$};
\end{tikzpicture}
\vskip 0.0mm
\caption{Contours of integrals \eqref{G} (upper contour) and \eqref{GG} (lower contour) and some of their poles. \label{figct}}
\end{figure}

This generalization is done by the technique of contour integral. Consider the following contour integrals, which can be easily shown to be zero since there is no pole inside the contours
\bea
&&\ointctrclockwise G(z) dz
=\ointctrclockwise \frac{1}{e^{-uzi}-1} \cdot \frac{ \sqrt{z^2+m^2} }{e^{\alpha\sqrt{z^2+m^2}}-1} dz
=0, \label{G}\\
&&\varointclockwise\tilde{G}(z) dz
=\varointclockwise \frac{1}{e^{uzi}-1} \cdot \frac{ \sqrt{z^2+m^2} }{e^{\alpha\sqrt{z^2+m^2}}-1} dz
=0, \label{GG}
\eea
where parameter~$u,~m,~\alpha$ are all positive.
The contours are shown in Fig. \ref{figct}, where $\rho$ is the radius of the small half or quarter circles. The width and height of each contour are specified by points $A$ and $B$ with values
\be \frac{2N\pi}{u}<A<\frac{2(N+1)\pi}{u}\mbox{ and } \sqrt{m^2+\frac{4N^2\pi^2}{\alpha^2}}<B<\sqrt{m^2+\frac{4(N+1)^2\pi^2}{\alpha^2}}\ee
where $N$ is some positive integer. In this plot, we also draw some of the poles of $G(z)$ that are relevant to the contour,
\be
z_{\rm left}=i\sqrt{m^2+\frac{4n^2\pi^2}{\alpha^2}},~z_{\rm down}=\frac{2n\pi}{u} \quad (n=1,~2,~\cdots,~N)
\ee
and poles of $\tilde{G}(z)$
\be
z'_{\rm left}=-i\sqrt{m^2+\frac{4n^2\pi^2}{\alpha^2}},~z_{\rm up}=\frac{2n\pi}{u}  \quad (n=1,~2,~\cdots,~N).
\ee

\bc
\ec

Equations \eqref{G} and \eqref{GG} lead to
\bea
&&\int_{\rho}^A\bigl[G(x)+\tilde{G}(x)\bigr]dx
+i\int_{0}^B\bigl[G(A+iy)-\tilde{G}(A-iy)\bigr]dy
+\int_{A}^0\bigl[G(x+iB)+\tilde{G}(x-iB)\bigr]dx\nonumber\\
&&+i\int_{B}^{\rho}\bigl[G(iy)-\tilde{G}(-iy)\bigr]dy
=i\pi \bigg\{ \frac{1}{2}\mbox{Res}~G(0)-\frac{1}{2}\mbox{Res}~\tilde{G}(0)
+\sum_{n=1}^N\Bigl[ \mbox{Res}~G(z_{\rm left})
+\mbox{Res}~G(z_{\rm down})\nonumber\\
&&-\mbox{Res}~\tilde{G}(z'_{\rm left})-\mbox{Res}~\tilde{G}(z_{\rm up}) \Bigr] \bigg\}, \label{G and GG}
\eea
where
\bea
&&\mbox{Res}~G(z_{\rm left})=-\mbox{Res}~\tilde{G}(z'_{\rm left}) =\frac{i4\pi^2n^2}{\alpha^3\sqrt{m^2+4n^2\pi^2/\alpha^2} (e^{u\sqrt{m^2+4n^2\pi^2/\alpha^2}}-1)}, \\
&&\mbox{Res}~G(z_{\rm down})=-\mbox{Res}~\tilde{G}(z_{\rm up}) =\frac{i\sqrt{m^2+4n^2\pi^2/u^2}/u}{e^{\alpha\sqrt{m^2+4n^2\pi^2/u^2}}-1},\\
&&\mbox{Res}~G(0)=-\mbox{Res}~\tilde{G}(0)=\frac{im}{u(e^{\alpha m}-1)}.
\eea
It is not hard to see that
\be
\lim\limits_{x{\rm ~or~}y\rightarrow\infty}G(x+iy)=
\lim\limits_{x{\rm ~or~}y\rightarrow\infty }\tilde{G}(x-iy)=0,
\ee
and the second and third integrals in Eq. \eqref{G and GG} vanish when $A$ and $B$, equivalently $N$, go to infinity. Then letting $u=2\pi/\theta$, Eq.~\eqref{G and GG} can be recast into the equality
\bea
\sum_{n} \frac{\sqrt{\theta^2 n^2+m^2}}{e^{\alpha\sqrt{\theta^2 n^2+m^2}}-1}
&=&-\frac{8\pi^3}{\theta\alpha^3}\sum_{n} \frac{n^2}{\sqrt{\frac{4\pi^2 n^2}{\alpha^2}+m^2}\Bigl(e^{\frac{2\pi}{\theta}\sqrt{\frac{4\pi^2 n^2}{\alpha^2}+m^2}}-1\Bigr)}
-\frac{m}{2(e^{\alpha m}-1)}\nonumber\\
&&+\frac{1}{\theta}\Biggl(\int_0^\infty\frac{\sqrt{x^2+m^2}}{e^{\alpha\sqrt{x^2+m^2}}-1}dx
+\int_m^\infty\frac{\sqrt{y^2-m^2}}{e^{\frac{2\pi}{\theta}y}-1}dy \Biggr). \label{one-sum}
\eea
Note that this equation implies Eq. \eqref{eq1}. This can be seen by setting $\theta=1$ and $m=0$ in Eq. \eqref{one-sum} and carrying out the integral using formula \cite{Wang}
\be
\int_0^\infty\frac{x^{s-1}e^{-ax}}{1-e^{-x}}dx=\Gamma(s)\zeta(s,a), \label{int of zeta}
\ee
where $\Gamma(s)$ is Gamma function and
\be
\zeta(s,a)=\sum_{n=0} \frac{1}{(n+a)^s}
\ee
is Hurwitz zeta function and $\zeta(s,1)\equiv \zeta(s)$ is the Riemann zeta function.

To generalize Eq. \eqref{one-sum} to the case of double series, we replace $m$ in it by $\sigma m$ and then sum over $m$. One then obtains
\bea
\sum_{m}\sum_{n}\frac{\sqrt{\theta^2 n^2+\sigma^2 m^2}}{e^{\sqrt{\theta^2 n^2+\sigma^2 m^2}}-1}&=&-\frac{8\pi^3}{\theta}\sum_{m}\sum_{n} \frac{n^2}{\sqrt{4\pi^2 n^2+\sigma^2 m^2}(e^{\frac{2\pi}{\theta}\sqrt{4\pi^2 n^2+\sigma^2 m^2}}-1)}
-\sum_{m}\frac{\sigma m}{2(e^{\sigma m}-1)}\nonumber\\
&&+\frac{1}{\theta}\sum_{m}\int_0^\infty\frac{\sqrt{x^2+\sigma^2 m^2}}{e^{\sqrt{x^2+\sigma^2 m^2}}-1}dx
+\frac{1}{\theta}\sum_{m}\int_{\sigma m}^\infty\frac{\sqrt{y^2-\sigma^2 m^2}}{e^{\frac{2\pi}{\theta}y}-1}dy, \label{two-expansion}
\eea
where $\theta,~\sigma>0$.
The first and second term on the right side will be kept. The third term can be calculated again using Eq.\eqref{one-sum}, and then for some terms using
\be
\frac{1}{e^y-1}=\sum_n e^{-yn}. \label{geometric series}
\ee
and lastly using
the definition of Bessel function of an imaginary argument~\cite{Gradshteyn}
\be
K_\nu(z)=\frac{(z/2)^\nu \Gamma(1/2)}{\Gamma(\nu+1/2)} \int_1^\infty e^{-zt}(t^2-1)^{\nu-1/2}dt \label{int of Bessel}
\ee
and the formula \eqref{int of zeta}. For the fourth term, we only need to use Eq. \eqref{geometric series} and \eqref{int of Bessel}. Combining all, final result of the double series Eq. \eqref{two-expansion} is given by
\bea
&&\sum_{m,n}\frac{\sqrt{\theta^2 n^2+\sigma^2 m^2}}{e^{\sqrt{\theta^2 n^2+\sigma^2 m^2}}-1}
=-\frac{8\pi^3}{\theta}\sum_{m,n} \frac{n^2}{\sqrt{4\pi^2 n^2+\sigma^2 m^2}(e^{\frac{2\pi}{\theta}\sqrt{4\pi^2 n^2+\sigma^2 m^2}}-1)}
-\frac{\sigma}{2}\sum_{m}\frac{m}{e^{\sigma m}-1}\nonumber\\
&&+\frac{1}{\theta} \biggl[-\frac{8\pi^3}{\sigma}Y_0\biggl(\frac{2\pi}{\sigma}\biggr) -\frac{\zeta(2)}{2}+\frac{\pi\zeta(3)}{\sigma}+\frac{\zeta(3)\sigma^2}{16\pi^2}
\biggr]
+\frac{\sigma}{2\pi}Y_1\biggl(\frac{\sigma}{\theta}\biggr), \label{two-sum}
\eea
where functions $Y_0(x)$ and $Y_1(x)$ are
\be
Y_0(x)=\sum_{m,n}m^2K_0(2\pi mnx), \quad Y_1(x)=\sum_{m,n}\frac{m}{n}K_1(2\pi mnx).\label{X0-Y1}
\ee

Eq.~\eqref{one-partition} can also be generalized to the cases of double series. Letting $\theta=\alpha/a$ and $\sigma=\alpha/b$, then dividing Eq.~\eqref{two-sum} by $\alpha$, indefinitely integrating both sides with respect to $\alpha$, and using the property of Bessel function \cite{Gradshteyn}
\be
\Bigl(\frac{d}{zdz} \Bigr)^i \bigl[z^\nu K_\nu(z)\bigr]=(-1)^i z^{\nu-i} K_{\nu-i}(z) \label{DK}
\ee
at $i=1$ for $Y_0$ yields
\bea
\sum_{n,m}\ln\Bigl(1-e^{-\alpha\sqrt{\frac{n^2}{a^2}+\frac{m^2}{b^2}}}\Bigr)
&=&Z_2\biggl(a,b,\frac{\alpha}{2\pi}\biggr)
-\frac{1}{2}Z_1\biggl(\frac{\alpha}{2\pi b}\biggr)
+a \biggl[\frac{\zeta(2)}{2\alpha}-\frac{\pi\zeta(3)b}{2\alpha^2}+\frac{\zeta(3)\alpha}{16\pi^2 b^2}
-\frac{2\pi}{\alpha}Y_1\biggl(\frac{2\pi b}{\alpha}\biggr) \biggr]\nonumber\\
&&+\frac{\alpha}{2\pi b}Y_1\biggl(\frac{a}{b}\biggr)+Q(a,b), \label{two-partition}
\eea
where $Z_1(x)$ and $Z_2(x,y,z)$ are
\be
Z_1(x)=\sum_m\ln\bigl(1-e^{-2\pi mx}\bigr), \quad Z_2(x,y,z)=\sum_{n,m}\ln\Bigl(1-e^{-2\pi x\sqrt{n^2/y^2+m^2/z^2}}\Bigr)\label{Z1-Z2}
\ee
and the integral constant $Q(a,b)$ is
\be
Q(a,b)
=\frac{1}{2}Z_1\biggl(\frac{a}{b}\biggr)
-\frac{\zeta(2)}{4\pi}+\frac{\zeta(3)b}{8\pi a}-\frac{\zeta(3)a^2}{8\pi b^2}
+Y_1\biggl(\frac{b}{a}\biggr)
-\frac{a}{b}Y_1\biggl(\frac{a}{b}\biggr) \label{Q(a,b)}.
\ee
Note that the $Z_1(\alpha/(2\pi b))$ in Eq.~\eqref{two-partition} can be calculated by Eq.~\eqref{one-partition}.

Finally, let us generalize Eq.~\eqref{eq1} and \eqref{one-partition} to the triple series case. Applying Eqs.~\eqref{one-sum},~\eqref{two-sum},~\eqref{geometric series} and~\eqref{int of Bessel} to the triple series,
and performing the summation in the order of $n$, $j$ and $m$, yields its result
\bea
&&\sum_{n,m,j}\frac{\sqrt{\theta^2 n^2+\sigma^2 m^2+\gamma^2 j^2}}{e^{\sqrt{\theta^2 n^2+\sigma^2 m^2+\gamma^2 j^2}}-1}
=-\frac{8\pi^3}{\theta}\sum_{k,m,j}\frac{k^2}{\sqrt{4\pi^2 k^2 + \sigma^2 m^2 + \gamma^2 j^2}(e^{\frac{2\pi}{\theta}\sqrt{4\pi^2 k^2+\sigma^2 m^2+\gamma^2 j^2}} - 1)}\nonumber\\
&&-\sum_{m,j}\frac{\sqrt{\sigma^2 m^2+\gamma^2 j^2}}{2(e^{\sqrt{\sigma^2 m^2+\gamma^2 j^2}}-1)}
+\frac{1}{\theta}
\biggl\{ \frac{-8\pi^3}{\gamma}\sum_{m,k,n}k^2 K_0\left(\frac{2\pi\sigma n}{\gamma}\sqrt{m^2+\frac{4k^2\pi^2}{\sigma^2}}\right)+\bigg[\frac{\zeta(2)}{4}%\\
-\frac{\pi\zeta(3)}{2\sigma}-\frac{\zeta(3)\sigma^2}{32\pi^2}+\frac{4\pi^3}{\sigma}X_0\biggl(\frac{2\pi}{\sigma}\biggr)\biggr]\nonumber\\
&&+\frac{1}{\gamma}\left[\frac{-\pi\zeta(3)}{2}+\frac{3\pi\zeta(4)}{\sigma}+\frac{\zeta(4)\sigma^3}{16\pi^3}%\\
-4\pi^4\left(\frac{2}{\pi\sigma}\right)^{\frac{1}{2}}\sum_{k,n}\left(\frac{k^5}{n}\right)^{\frac{1}{2}}K_{\frac{1}{2}}\biggl(\frac{4\pi^2 kn}{\sigma}\biggr)\right]
+\frac{\gamma^{\frac{1}{2}}\sigma^{\frac{3}{2}}}{4\pi}Y_{\frac{3}{2}}\biggl(\frac{\sigma}{\gamma}\biggr)\biggr\}%\\
+\frac{1}{2\pi}V_1\biggl(\frac{1}{\theta},\frac{1}{\sigma},\frac{1}{\gamma}\biggr) \label{three-sum}
\eea
where variables $\theta,~\sigma,~\gamma>0$, and $Y_{3/2}(x)$, $V_1(x,y,z)$ are
\be
Y_{\frac{3}{2}}(x)=\sum_{m,n}\biggl(\frac{m}{n}\biggr)^{\frac{3}{2}}K_{\frac{3}{2}}(2\pi mnx), \quad  V_1(x,y,z)=\sum_{k,m,n}\frac{\sqrt{m^2/y^2+k^2/z^2}}{n} K_1\biggl(2\pi nx\sqrt{m^2/y^2+k^2/z^2}\biggr).\label{Y3/2-V1}
\ee
Furthermore, in Eq. \eqref{three-sum} letting $\theta=\alpha/a$, $\sigma=\alpha/b$ and $\gamma=\alpha/c$, then dividing by $\alpha$ and indefinitely integrating both sides with respect to $\alpha$ yields
%three index partition
\bea
&&\sum_{n,m,j}\ln\Bigl(1-e^{-\alpha\sqrt{\frac{n^2}{a^2}+\frac{m^2}{b^2}+\frac{j^2}{c^2}}}\Bigr)
=Z_3\biggl(a,b,c,\frac{\alpha}{2\pi}\biggr)
-\frac{1}{2}Z_2\biggl(\frac{\alpha}{2\pi},b,c\biggr)
+a \biggl\{-V_1\biggl(c,b,\frac{\alpha}{2\pi}\biggr)\nonumber\\
&&+\biggl[\frac{-\zeta(2)}{4\alpha}+\frac{\pi\zeta(3)b}{4\alpha^2}-\frac{\zeta(3)\alpha}{32\pi^2 b^2}
+\frac{\pi}{\alpha}Y_1\biggl(\frac{2\pi b}{\alpha}\biggr) \biggr]
+c\biggl[\frac{\pi\zeta(3)}{4\alpha^2}-\frac{\pi\zeta(4)b}{\alpha^3}+\frac{\zeta(4)\alpha}{16\pi^3 b^3}
-\pi(\frac{2\pi}{b\alpha^3})^{1/2}Y_{\frac{3}{2}}\biggl(\frac{2\pi b}{\alpha}\biggr) \biggr]\nonumber\\
&&+\frac{\alpha}{4\pi c^{1/2}b^{3/2}}Y_{\frac{3}{2}}\biggl(\frac{c}{b}\biggr) \biggr\}
+\frac{\alpha}{2\pi}V_1(a,b,c)
+N(a,b,c), \label{three-partition}
\eea
where $Z_2(x,y,z)$ was defined in Eq. \eqref{Z1-Z2}, and function $Z_3(x,y,z,t)$ and the integral constant $N(a,b,c)$ are respectively
\bea
Z_3(x,y,z,t)&=&\sum_{n,m,j}\ln\bigl(1-e^{-2\pi x\sqrt{n^2/y^2+m^2/z^2+j^2/t^2}}\bigr),\label{Z3} \\
N(a,b,c)
&=&\frac{1}{2}Z_2(a,b,c)
+a\biggl[ V_1(c,a,b)%\\
-V_1(a,b,c)\biggr]
+\frac{\zeta(2)}{8\pi}-\frac{1}{2}Y_1\biggl(\frac{b}{a}\biggr)
+\biggl[\frac{\zeta(3)a^2}{16\pi b^2}-\frac{\zeta(3)b}{16\pi a}-\frac{\zeta(3)c}{16\pi a}\biggr]\nonumber\\
&&+c\biggl[\frac{\zeta(4)b}{8\pi^2 a^2}-\frac{\zeta(4)a^2}{8\pi^2b^3}\biggr]
+\frac{c}{2b^{1/2}a^{1/2}}Y_{\frac{3}{2}}\biggl(\frac{b}{a}\biggr)
-\frac{a^2}{2c^{1/2}b^{3/2}}Y_{\frac{3}{2}}\biggl(\frac{c}{b}\biggr). \label{W(a,b,l)}
\eea
Eqs. \eqref{three-sum} and \eqref{three-partition} are the generalizations of Eqs. \eqref{eq1} and \eqref{one-partition} to the triple series case. In addition, when the sign in the $\ln$ function in Eq. \eqref{three-partition} is changed from minus to plus, one can reach the formula
\bea
&&\sum_{n,m,j}\ln\Bigl(1+e^{-\alpha\sqrt{\frac{n^2}{a^2}+\frac{m^2}{b^2}+\frac{j^2}{c^2}}}\Bigr)
=Z_3\biggl(\frac{\alpha}{\pi},a,b,c\biggr)-Z_3\biggl(\frac{\alpha}{2\pi},a,b,c\biggr)\nonumber\\
&&=\frac{7\pi\zeta(4)abc}{8\alpha^3}
+\alpha\biggl[\frac{\zeta(4)ac}{16\pi^3b^3}
+\frac{a}{4\pi b^{3/2} c^{1/2}}Y_{\frac{3}{2}}\biggl(\frac{c}{b}\biggr)
+\frac{1}{2\pi}V_1(a,b,c) \biggr]
+Z_3\biggl(a,b,c,\frac{\alpha}{\pi}\biggr)-Z_3\biggl(a,b,c,\frac{\alpha}{2\pi}\biggr)\nonumber\\
&&+aV_1\biggl(c,b,\frac{\alpha}{2\pi}\biggr)-aV_1\biggl(c,b,\frac{\alpha}{\pi}\biggr)
+ac\pi\biggl(\frac{2\pi}{b\alpha^3}\biggr)^{1/2}Y_{\frac{3}{2}}\biggl(\frac{2\pi b}{\alpha}\biggr)-\frac{ac\pi}{2}\biggl(\frac{\pi}{b\alpha^3}\biggr)^{1/2}Y_{\frac{3}{2}}\biggl(\frac{\pi b}{\alpha}\biggr) \nonumber \\
&&+\frac{\zeta(2)a}{8\alpha}
-\frac{3\pi\zeta(3)ab}{16\alpha^2}
-\frac{3\pi\zeta(3)ac}{16\alpha^2}
-\frac{\zeta(3)a\alpha}{32\pi^2b^2}
+\frac{1}{2}Z_2\biggl(\frac{\alpha}{2\pi},b,c\biggr)-\frac{1}{2}Z_2\biggl(\frac{\alpha}{\pi},b,c\biggr)
+\frac{a\pi}{2\alpha}Y_1\biggl(\frac{\pi b}{\alpha}\biggr)-\frac{a\pi}{\alpha}Y_1\biggl(\frac{2\pi b}{\alpha}\biggr). \label{three partition plus}
\eea
which will be useful for the calculation of fermionic field Casimir effect.

\section{Thermal Casimir Effect of Electromagnetic field in a rectangular box\label{EM field}}
Geyer et al.~\cite{Geyer:2008wb} studied the Casimir effect of electromagnetic field in ideal metal rectangular boxes at finite temperature. They used Abel-Plana formula to calculate the non-renormalized thermal correction term $\Delta_TF_0$ in the renormalized free energy $F^{\rm phys}$ of the electromagnetic field. In this section, we calculate $F^{\rm{phys}}$ using the generalized  Schl\"omilch's formula developed in Eq. \eqref{three-partition} for arbitrary edge sizes and temperature.

The renormalized free energy of electromagnetic field confined in a three-dimensional box is given by \cite{Geyer:2008wb}
\be
F^{\rm{phys}}(a,b,c,T)=E_0^{\rm{ren}}(a,b,c)+\Delta_T F_0(a,b,c,T)-F_{\rm bb}(a,b,c,T)-\alpha_1^{el}T^3-\alpha_2^{el}T^2, \label{FFE}
\ee
where $a,~b,~c$ are the edge sizes of the box, $T$ is the temperature and $E_0^{\rm{ren}}(a,b,c)$ is the renormalized free energy at zero temperature. $\Delta_T F_0(a,b,c,T)$ is the non-renormalized thermal correction
\be
\Delta_T F_0(a,b,c,T)=T\biggl[ \sum_{n,m}\ln(1-e^{-\frac{\omega_{nm0}}{T}})
+\sum_{n,j}\ln(1-e^{-\frac{\omega_{n0j}}{T}})
+\sum_{m,j}\ln(1-e^{-\frac{\omega_{0mj}}{T}})
+2\sum_{n,m,j}\ln(1-e^{-\frac{\omega_{nmj}}{T}}) \biggr]. \label{logarithm partition}
\ee
where
\be
\omega_{nmj}=\pi\sqrt{\frac{n^2}{a^2}+\frac{m^2}{b^2}+\frac{j^2}{c^2}},\quad n,m,j=1,2,...
\ee
are frequencies, and
\be
F_{\rm bb}(a,b,c,T)=-\frac{\pi^2 T^4 abc}{45} \label{Fbb}
\ee
is the free energy of the black body radiation. Finally, $\alpha_1^{el}$ and $\alpha_2^{el}$ are coefficients of two renormalization terms which should cancel the corresponding terms in $\Delta_T F_0(a,b,c,T)$ to prevent possible high temperature divergence that can contribute to the Casimir force.

The renormalized free energy at zero temperature $E_0^{\rm{ren}}(a,b,c)$ can be calculated using the Abel-Plana formula \cite{Mostepanenko:1997sw, Bordag:2001qi} and Epstein zeta function \cite{Bordag:2001qi}. Starting from the definition of the non-renormalized zero temperature free energy $E_0(a,b,c)$
\be
E_0(a,b,c)=\frac{1}{2}\biggl( 2\sum_{n,m,j}\omega_{nmj}+\sum_{n,m}\omega_{nm}+\sum_{n,j}\omega_{nj}+\sum_{m,j}\omega_{mj} \biggr), \label{infite}
\ee
one can use the Abel-Plana formula~\cite{Bordag:2001qi}
\be
\sum_{n=0} g(n)
-\int_0^\infty g(t)dt
=\frac{g(0)}{2}
+i\int_0^\infty\frac{g(it)-g(-it)}{e^{2\pi t}-1}dt. \label{Abel-Plana formula}
\ee
where $g(z)$ is any analytic function in the right half-plane to perform the summation in Eq.~\eqref{infite} in the order of $n$, $j$ and $m$. This allows us to separate its infinite parts
to obtain the finite renormalized free energy at zero temperature $E_0^{\rm{ren}}(a,b,c)$ as
\be
E_0^{\rm{ren}}(a,b,c)=-a\biggl[ \frac{\zeta(4)c}{8\pi^2 b^3}+\frac{\zeta(3)}{16\pi c^2}+\frac{1}{2b^{3/2}c^{1/2}}Y_{\frac{3}{2}}\biggl(\frac{c}{b}\biggr)\biggr]
+\frac{\pi}{48}\biggl(\frac{1}{b}+\frac{1}{c} \biggr)
-\biggl[V_1(a,b,c)
+\frac{1}{2b}Y_1\biggl(\frac{a}{b}\biggr)
+\frac{1}{2c}Y_1\biggl(\frac{a}{c}\biggr)\biggr]. \label{E0ren}
\ee
where $Y_{3/2}(x)$ and $V_1(x,y,z)$ were defined in Eq.~\eqref{Y3/2-V1}, and $Y_1(x)$ was defined in Eq.~\eqref{X0-Y1}.

For the computation of the non-renormalized thermal correction $\Delta_TF_0(a,b,c,T)$, our approach is different from the Abel-Plana formula method used by Ref.~\cite{Geyer:2008wb}. Instead, in this paper we calculate it using the generalized Schl\"omilch's formula obtained in Sec.~\ref{series}. Applying Eqs.~\eqref{three-partition},\eqref{two-partition} and \eqref{one-partition} to Eq. \eqref{logarithm partition}, the analytical form of $\Delta_T F_0(a,b,c,T)$ can be obtained as
%bosonic free energy
\be
\Delta_TF_0(a,b,c,T)=-\frac{T\ln T}{2}
+TF_1(a,b,c)+F_2(a,b,c,T)-E_0^{\rm{ren}}(a,b,c)
-\frac{2\zeta(4)abc}{\pi^2}T^4+\frac{\pi(a+b+c)}{12}T^2, \label{bosonic free energy}
\ee
where
\bea
F_1(a,b,c)&=&\frac{\zeta(4)bc}{4\pi^2 a^2} -a^2\biggl[ \frac{\zeta(4)c}{4\pi^2 b^3}+\frac{\zeta(3)}{8\pi c^2}+\frac{1}{b^{3/2}c^{1/2}}Y_{\frac{3}{2}}\biggl(\frac{c}{b}\biggr)\biggr]%\\
+\biggl[Z_2(a,b,c)
+\frac{1}{2}Z_1\biggl(\frac{a}{b}\biggr)
+\frac{1}{2}Z_1\biggl(\frac{a}{c}\biggr)
+Y_1\biggl(\frac{c}{a}\biggr)\biggr]\nonumber\\
&&-\biggl[2aV_1(a,b,c)
+\frac{a}{b}Y_1\biggl(\frac{a}{b}\biggr)
+\frac{a}{c}Y_1\biggl(\frac{a}{c}\biggr)\biggr]
+2aV_1(c,b,a)
+\frac{c}{\sqrt{ab}}Y_{\frac{3}{2}}\biggl(\frac{b}{a}\biggr)%\\
-\frac{\ln(bc)}{4}-\frac{\zeta(2)}{4\pi}-\frac{\ln 2}{2}, \label{F1(abl)}\\
F_2(a,b,c,T)&=&T\biggl\{
\biggl[ 2Z_3\biggl(a,b,c,\frac{1}{2T}\biggr)
        +Z_2\biggl(a,b,\frac{1}{2T}\biggr)%\\
        +Z_2\biggl(a,c,\frac{1}{2T}\biggr)
        -\frac{1}{2}Z_1(2Tb)
        -\frac{1}{2}Z_1(2Tc) \biggr]\nonumber\\
&&-a\biggl[2V_1\biggl(c,b,\frac{1}{2T}\biggr)
         +2c\biggl(\frac{2T^3}{b}\biggr)^{1/2}Y_{\frac{3}{2}}(2bT)%\\
         +2TY_1(2cT) \biggr]  \biggr\},  \label{F2(abl-alpha)}.
\eea
Equation~\eqref{bosonic free energy} implies that in Eq. \eqref{FFE}
\be
\alpha_1^{el}=0, \quad \alpha_2^{el}=\pi(a+b+c)/12. \label{alpha-1-2}
\ee
Eqs. \eqref{alpha-1-2} agrees with Ref.~\cite{Geyer:2008wb} which computed the high temperature limit of the Casimir energy.

Substituting Eq. \eqref{bosonic free energy} into the renormalized free energy Eq. \eqref{FFE}, the final Casimir energy of electromagnetic field in a three-dimensional rectangular box at finite temperature is finally written as
\be
F^{\rm{phys}}(a,b,c,T)=-\frac{T\ln T}{2}+TF_1(a,b,c)+F_2(a,b,c,T). \label{E physic}
\ee

In order to compare with previous works, we computed the high and low temperature limits of \eqref{E physic}. At high temperature, we can show in Appendix \ref{Appendix2} that the last term $F_2(a,b,c,T)$ in Eq.~\eqref{E physic} approaches zero. Therefore the Casimir energy becomes
\be
F^{\rm{phys}}(a,b,c,T\to\infty)=-(T\ln T)/2+TF_1(a,b,c). \label{E physic_ht_old}
\ee
The first term here is geometry independent and therefore dose not contribute to the electromagnetic Casimir force. Moreover, because it is negative and divergent at infinite temperature, this term should be subtracted in order to get a physically meaningful Casimir energy. Finally we have
\be F^{\rm{phys}}(a,b,c,T\to\infty)= TF_1(a,b,c). \label{E physic_ht}\ee
This shows that at high temperature, the temperature dependence of the Casimir energy is particularly simple, while the edge size dependence is solely through the term $F_1(a,b,c)$.

In the low temperature limit, we can show that the entire $\Delta_TF_0(a,b,c,T)$  in Eq.~\eqref{logarithm partition} goes zero (see the steps from Eq.~\eqref{lim 1} to Eq.~\eqref{inequlity 1}). Therefore using definition \eqref{FFE},  the renormalized free energy in low temperature becomes $E_0^{\rm{ren}}(a,b,c)$ given in Eq.~\eqref{E0ren}
\be F^{\rm{phys}}(a,b,c,T\to 0)=E_0^{\rm{ren}}(a,b,c). \ee
If one is interested in the electromagnetic Casimir energy of a cube at zero temperature, then setting $a=b=c$ in Eq. \eqref{E0ren} produces numerically
\be
F^{\rm{phys}}(a,a,a,T\to 0)=E_0^{\rm{ren}}(a,a,a)=\frac{0.0917}{a}, \label{F0aaa}
\ee
and therefore an attractive force between opposite faces of the cube. Eq. \eqref{F0aaa} agrees well with the result obtained in Eq. (72) of Ref. \cite{Geyer:2008wb}.

\section{Thermal Casimir energy and force of fermionic field in a rectangular box\label{Fermion field}}
In this section the Casimir effect at finite temperature for a massless fermionic field confined in a three-dimensional box will be calculated using the M.I.T. bag model boundary condition. This condition allows no flux through the boundary and leads to the discrete momenta of the form~\cite{Seyedzahedi:2010fya}
\be
p_i=\left(\frac{1}{2}+n_i\right)\frac{\pi}{l_i},\quad l_1=a,~l_2=b,~l_3=c, \quad n_i=0,1,2,\cdots. \label{momenta}
\ee
The non-renormalized free energy for the field is defined as
\be
F=F_{\rm b0}+F_T=4\sum_{n,m,j=0}\Bigl(-\frac{1}{2}\omega_{nmj}\Bigr)-4T\sum_{n,m,j=0}\ln\Bigl(1+e^{-\frac{\omega_{nmj}}{T} }\Bigr), \label{infinite fermionic energy}
\ee
where the first term $F_{\rm b0}$ is non-renormalized energy at zero temperature, the second term $F_T$ is the non-renormalized thermal correction, and
\be
\omega_{nmj}=\sqrt{p_1^2+p_2^2+p_3^2}=\pi\sqrt{\frac{(n+1/2)^2}{a^2}+\frac{(m+1/2)^2}{b^2}+\frac{(j+1/2)^2}{c^2}}
\ee
are frequencies. The factor 4 appears in Eq.~\eqref{infinite fermionic energy} because of the antiparticle and spin multiplicities~\cite{Johnson}. In the following, we will compute these two terms one by one using formulas obtained in section \ref{series}.

The non-renormalized energy at zero temperature had been calculated by Seyedzahedi et al.~\cite{Seyedzahedi:2010fya} for a cube by using a modified form of the Abel-Plana formula~\cite{Bordag:2001qi,Saharian:2007ph}
\be
\sum_{n=0} g\Bigl(n+\frac{1}{2}\Bigr)
=\int_0^\infty g(t)dt
-i\int_0^\infty\frac{g(it)-g(-it)}{e^{2\pi t}+1}dt. \label{Mofified Abel-Plana formula}
\ee
where $g(z)$ is analytic in the right half-plane. For arbitrary edge sizes, we can also use this formula to Eq.~\eqref{infinite fermionic energy},  by first performing the summation for $n$, then for $j$ and eventually for $m$. Eventually one finds for the $F_{\rm b0}$ term
\be
F_{\rm b0}=F_0(a,b,c)+F_{\rm f0}(a,b,c), \label{Fb0}
\ee
where
\be
F_0(a,b,c)=-\biggl\{ \frac{7\zeta(4)ac}{32\pi^2 b^3}
+2M_1(a,b,c) +\frac{a}{b^{3/2}c^{1/2}}M_{\frac{3}{2}}\biggl(\frac{c}{b}\biggr) \biggr\} \label{F-0}
\ee
is the renormalized energy at zero temperature.  Here functions $M_{3/2}(x)$ and $M_1(x,y,z)$ are defined as
\bea
M_1(x,y,z)&=&\sum_{m,k=0}\sum_{n=1} \frac{(-1)^{n+1}}{n}\sqrt{\frac{(m+1/2)^2}{y^2}+\frac{(k+1/2)^2}{z^2}} K_1 \biggl(2\pi xn\sqrt{\frac{(m+1/2)^2}{y^2}+\frac{(k+1/2)^2}{z^2}}\biggr), \label{M1}\\
M_{\frac{3}{2}}(x)&=&\sum_{m=0}\sum_{n=1} \frac{(-1)^{n+1}}{n^{3/2}}\left(m+\frac{1}{2}\right)^{3/2}K_{\frac{3}{2}}\bigl(2\pi(m+1/2)nx\bigr). \label{M3/2}
\eea
Moreover,
\be
F_{\rm f0}(a,b,c)=-2\pi\int_0^\infty dx\int_0^\infty dy\int_0^\infty dz \sqrt{\frac{x^2}{a^2}+\frac{y^2}{b^2}+\frac{z^2}{c^2}} \label{Ff0}
\ee
is the energy at zero temperature in the absence of the boundaries, which should be subtracted later.

The second term $F_T$ in Eq.~\eqref{infinite fermionic energy}, is where our result in Sec. \ref{series}, i.e., Eq. \eqref{three partition plus} will be used. As will be shown later, it is through the usage of this equation that the blackbody radiation term in the free energy can be subtracted from the non-renormalized energy to obtain a meaningful Casimir energy. Equation \eqref{three partition plus} after some tedious algebra (see Appendix \ref{appd1}) yields the final result for $F_T$
\be
F_T=4TA_3(a,b,c,T)+F_{\rm fb}(a,b,c,T)-F_0(a,b,c), \label{FT}
\ee
where $F_0$ is the same as in Eq. \eqref{F-0} and
\be
A_3(a,b,c,T)=-W_3\biggl(a,b,c,\frac{1}{2T}\biggr)
-aM_1\biggl(c,b,\frac{1}{2T}\biggr)
-ac\biggl(\frac{2T^3}{b}\biggr)^{1/2}M_{\frac{3}{2}}(2bT) .\label{A3}
\ee
Here $M_1(x,y,z)$ and $M_{3/2}(x)$ were defined in Eqs. \eqref{M1} and \eqref{M3/2} and $W_3(x,y,z,t)$ is
\be
W_3(x,y,z,t)=\sum_{m,n,k=0}\ln\biggl(1+e^{-2\pi x\sqrt{\frac{(m+1/2)^2}{y^2}+\frac{(n+1/2)^2}{z^2}+\frac{(k+1/2)^2}{t^2}}}\biggr). \label{W3}
\ee
The second term $F_{\rm fb}$ on the right side of Eq.~\eqref{FT} is found to be
\be
F_{\rm fb}(a,b,c,T)=-\frac{7\zeta(4)abc}{2\pi^2}T^4. \label{Ffb}
\ee
It is easy to see that this term is indeed the free black body radiation energy, namely the free energy at finite temperature in the absence of boundaries
\be
-4T\int_{-\infty}^\infty \frac{d^3p}{(2\pi)^3}\ln\bigl(1+e^{-\frac{\omega_p}{T}}\bigr)\cdot abc=-\frac{7\zeta(4)abc}{2\pi^2}T^4=F_{\rm fb}(a,b,c,T). \label{Fu}
\ee

Substituting Eq.~\eqref{Fb0} and Eq.~\eqref{FT} into Eq.~\eqref{infinite fermionic energy} yields the non-renormalized free energy
\be
F=4T A_3(a,b,c,T)+F_{\rm f0}(a,b,c)+F_{\rm fb}(a,b,c,T). \label{F}
\ee
To obtain the Casimir energy, the free energy in the absence of boundaries, namely the last two terms in Eq.~\eqref{F}, should be subtracted from $F$. Thus the final renormalized free Casimir energy is
\be
F_{\rm{C}}=4T A_3(a,b,c,T). \label{FC}
\ee
As mentioned previously, it is seen here that the removal of the thermal contribution $F_{\rm fb}(a,b,c,T)$ to the non-renormalized energy is done by computation using Eq. \eqref{three partition plus} and correctly recognize the continuous black body radiation term $F_{\rm fb}(a,b,c,T)$.

It is also clear that the above Casimir free energy  $F_{\rm{C}}$ will not depend on the order of edges $a,~b,~c$ but only their sizes because the same set of $\{a,~b,~c\}$ always define a fixed rectangular box. We can calculate the Casimir force between any pair of opposite faces of the box. Here we choose the pair perpendicular to edges $a$ and then taking derivative with respect to $a$ produces the Casimir force
\be
f_a=-\frac{\partial F_{\rm{C}}}{\partial a}
=4T\biggl\{
 -2\pi\sum_{k,m,j=0} \frac{\sqrt{4T^2(k+1/2)^2+\frac{(m+1/2)^2}{b^2}+\frac{(j+1/2)^2}{c^2}}}{e^{2\pi a\sqrt{4T^2(k+1/2)^2+\frac{(m+1/2)^2}{b^2}+\frac{(j+1/2)^2}{c^2}}}+1}
+M_1\biggl(c,b,\frac{1}{2T}\biggr)
+c\biggl(\frac{2T^3}{b}\biggr)^{1/2}M_{\frac{3}{2}}(2bT) \biggr\}. \label{fa}
\ee
Let us emphasis that these are general formulas, i.e., Eqs. \eqref{FC} and \eqref{fa}, valid for any values of lengths $a,~b,~c$ and temperature $T$ is obtained for a massless fermionic field in a three-dimensional rectangular box.

\section{Effects of temperature and edge lengths in fermionic Casimir effect\label{secparaspace}}

With the full result of the Casimir energy \eqref{FC} and force \eqref{fa} for a fermionic field in a  rectangular box with  arbitrary sizes $(a,~b,~c)$ and temperature $T$, we can now do a full analysis of these two quantities in the entire parameter space spanned by these four parameters.

First of all, we can reduce the full parameter space $a\in(0,\infty)\times b\in(0,\infty)\times c\in(0,\infty)\times T\in(0,\infty)$ into a smaller one by taking advantage of the cyclic symmetry of the sizes $(a,~b,~c)$. That is, we will assume $b\leq c\leq a$ without losing any generality. This effectively reduce the parameter space to one-eighth of the original one. Moreover, since $1/T$ has the same dimension as length in our convention of units ($\hbar=c=\kappa_B=1$), we can directly compare it with the edge lengths. With these simplifications, we will be able to do a full analysis of the Casimir energy and force in the reduced parameter space. We will study in turn the low and high temperature limits, and then the finite temperature case. In each case, we scan some ranges of the  parameters and look for interesting features of the Casimir energy and force.

\subsection{Low temperature limit: $1/T\to\infty$ \label{casea}}
In this limit, because the only dimensional variable of the inputs are $a,~b$ and $c$, the Casimir energy will depend only on one absolute scale, for which we chose $b$, and then the ratios between the edges. This not only means that the effective parameter space is further reduced, but also that the Casimir energy will take the form
\be
F_{\rm{C}}=\frac{1}{b}g\left(\frac{a}{b},~\frac{c}{b}\right),\ee
where $g$ is some function depending on $a/b$ and $c/b$ only. This indeed can be simply verified from Eq. \eqref{FC}. Therefore, without losing any generality, we can set $b=1$. There will exist three subcases: (A1) all of the edge sizes $b,~c,~a$ are finite, i.e., a three-dimensional box; (A2) $b, ~c$ are finite and $a$ is infinite, i.e., a waveguide and (A3) $b=1$ is finite and $c,~a$ are infinite, i.e., two parallel plates.

We can simply compute the zero temperature limit of the Casimir energy and Casimir force for arbitrary edge sizes.
According to Eq.~\eqref{infinite fermionic energy}, $F_T$ approaches zero when $T$ goes to zero. Hence, at zero temperature the Casimir energy \eqref{FC} turns into the renormalized energy $F_0$ in Eq.~\eqref{F-0}
\be
F_C(a,b,c,T\to 0)=F_0(a,b,c)=-\biggl\{ \frac{7\zeta(4)ac}{32\pi^2 b^3}
+\frac{a}{b^{3/2}c^{1/2}}M_{\frac{3}{2}}\biggl(\frac{c}{b}\biggr) +2M_1(a,b,c) \biggr\}
\label{t0ce}
\ee
which is the same as \eqref{F-0} and by using formula \eqref{DK} the Casimir force is given by
\be
f_{\rm 0a}(a,b,c)=-\frac{\partial}{\partial a}F_0(a,b,c) =\frac{7\zeta(4)c}{32\pi^2 b^3}+\frac{1}{b^{3/2}c^{1/2}}M_{\frac{3}{2}}\biggl(\frac{c}{b}\biggr)
-\frac{2}{a}M_1(a,b,c)-4\pi M_0(a,b,c),
\label{t0cf}
\ee
where $M_{3/2}(x)$ and $M_1(x,y,z)$  were defined in Eqs. \eqref{M3/2} and \eqref{M1} and
\be
M_0(x,y,z)=\sum_{m,k=0}\sum_{n=1}(-1)^{n+1}\biggl[\frac{(m+\frac{1}{2})^2}{y^2}+\frac{(k+\frac{1}{2})^2}{z^2}\biggr]K_0\biggl(2\pi nx\sqrt{\frac{(m+\frac{1}{2})^2}{y^2}+\frac{(k+\frac{1}{2})^2}{z^2}}\biggr). \label{M0}
\ee
When $b=c=a$, these equations yield the Casimir energy and Casimir force of a cube
\bea
F_0(a,a,a)=-\frac{1}{a}\Bigl(\frac{7\pi^2}{2880}+0.0142+0.0108\Bigr)=-0.0489\frac{1}{a},\label{F0aaafermion}\\
f_{\rm 0a}(a,a,a)=\frac{1}{a^2}\Bigl(\frac{7\pi^2}{2880}+0.0142-0.0108-0.0437\Bigr)=-0.0163\frac{1}{a^2}. \label{f0aaa}
\eea
Our results in this special case agrees perfectly with previous calculation in Eqs. (A16) of Ref. \cite{Seyedzahedi:2010fya} done for this geometry.

\begin{figure}[htp]
(a)\hspace{7cm} (b)\\
\vspace{5mm}
\includegraphics[scale=0.4]{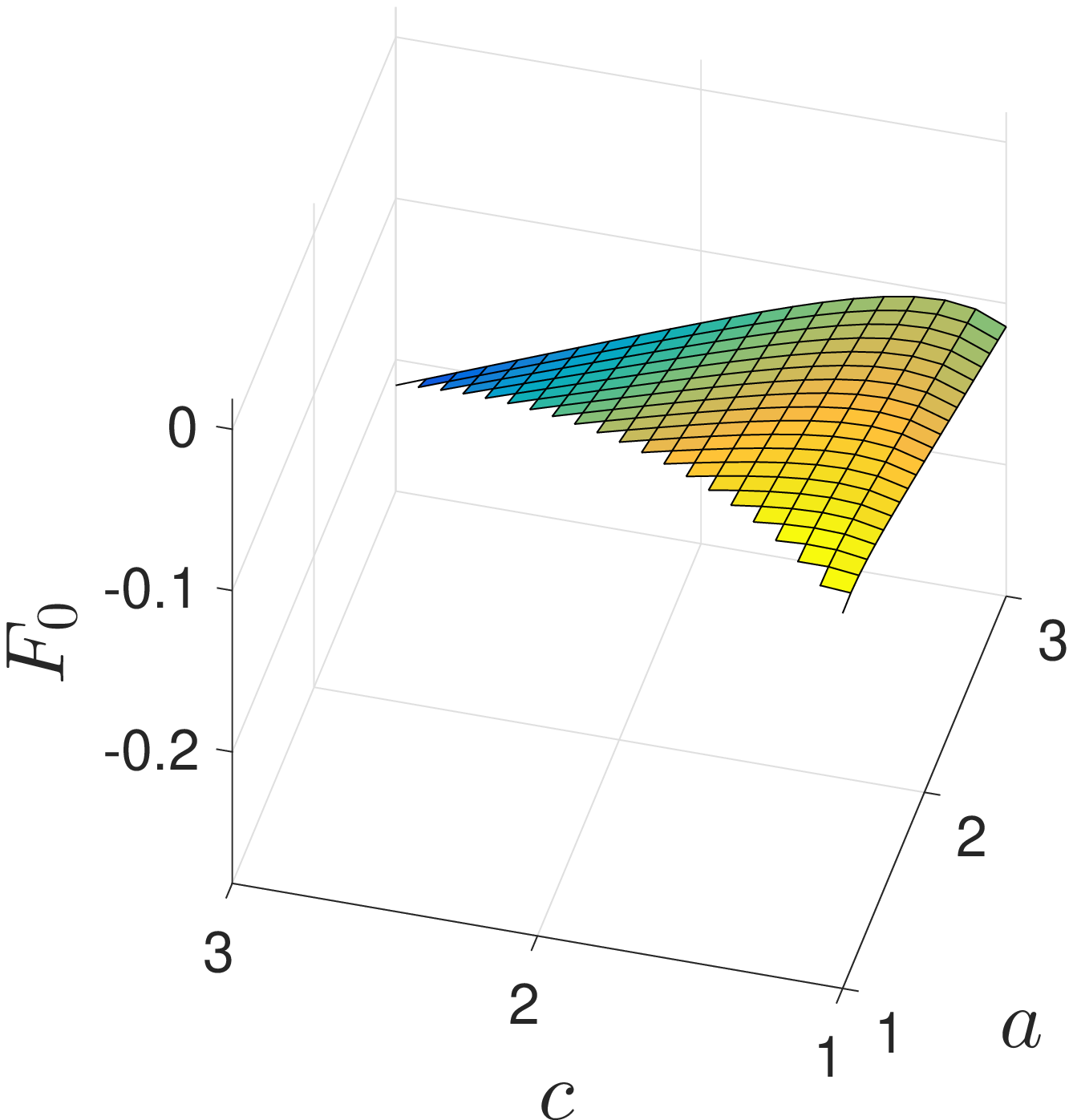}\hspace{4mm}
\includegraphics[scale=0.4]{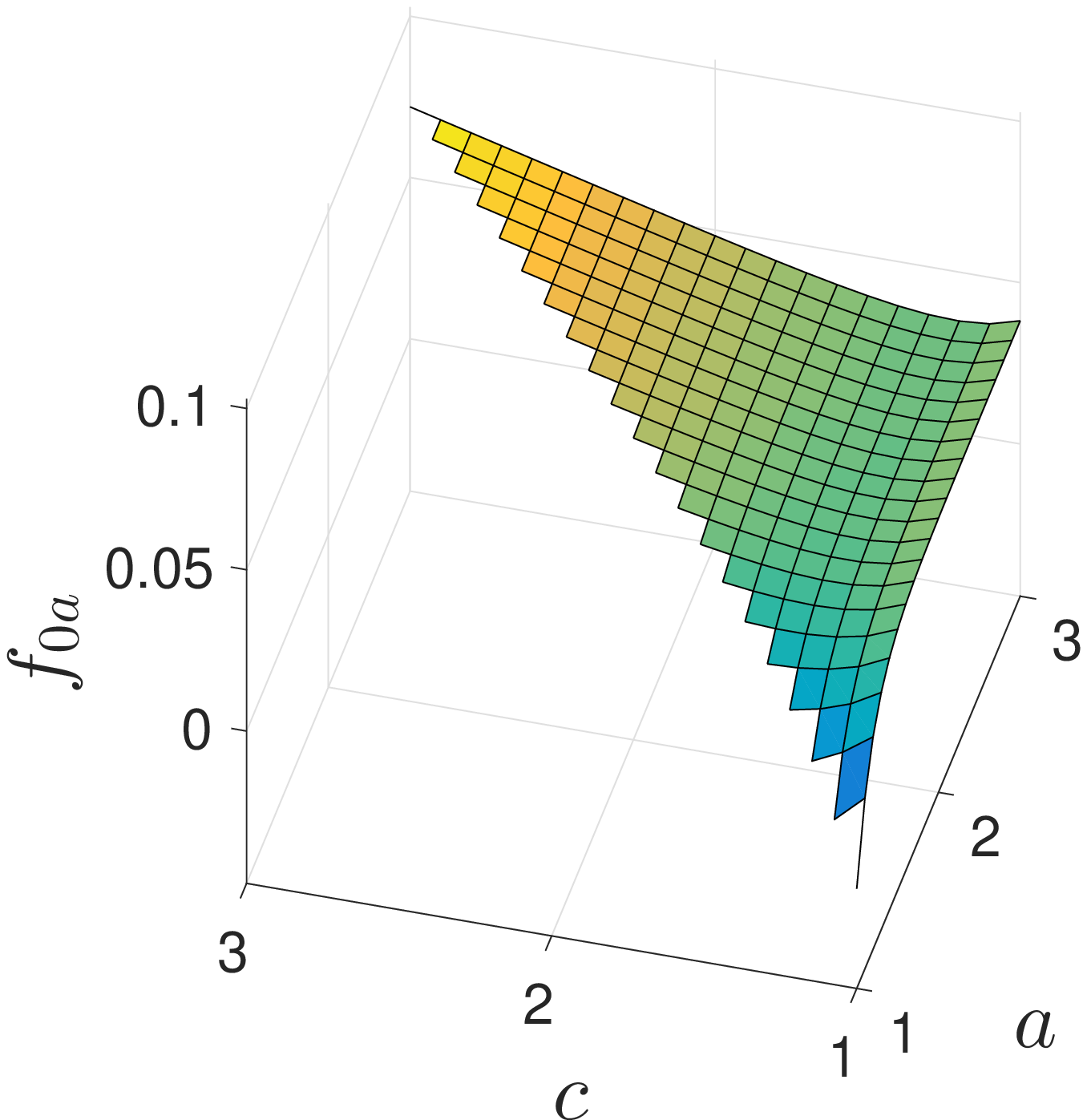}\\
(c)\hspace{7cm} (d)\\
\vspace{5mm}
\includegraphics[scale=0.4]{fig2t0cfrecc.eps}\hspace{4mm}
\includegraphics[scale=0.4]{fig2fb.eps}\\
(e)\\
\vspace{5mm}
\includegraphics[scale=0.4]{fig2contournew3.eps}
\caption{(a): Casimir energy for a fermionic field in a rectangular box at zero temperature. (b),(c) and (d): the corresponding Casimir force along $b$, $c$ and $a$ directions respectively. (e) The force transition boundaries.  Choice of parameters are $b=1$, $c$ from 1 to 3 and $a$ from $c$ to 3. \label{t0fig1}}
\end{figure}

For case (A1), we studied numerically the Casimir energy and force for a range of parameters  using the above formulas. We plotted in Fig. \ref{t0fig1} the Casimir energy for $b=1$ and $c$ from $1$ to 3 and $a$ from $c$ to 3 and the corresponding Casimir forces along $a$ and $c$ directions respectively.

It is seen that the Casimir energy is always negative while the sign of the forces in neither the $a$ nor the $c$ direction is fixed. The magnitude of the force in the $a$ direction is in general smaller than that in the $c$ direction, which is understandable because $a>c$ in this part of the parameter space. Also because of this, the forces in both the $a$ and $c$ direction change much slower as $a$ varies than they change as $c$ varies. For the force in the $a$ direction, as one can see from Fig. \ref{t0fig1}(b), it will change from repulsive to attractive as $a$ decreases to almost one while $c$ was kept a small constant $c\sim 1$. For the force in the $c$ direction, Fig. \ref{t0fig1}(c) shows that the force also transforms from repulsive to attractive, but mainly with the decrease of $c$ from much larger than 1 to about 1.  These changes of sign of the force was also reported in Ref. \cite{Gundersen:1987wz} for parallel plates and in Ref \cite{Queiroz:2004wi} for three-dimensional box.
Lastly, the force in the $b$ direction is independent from the forces along the $a$ and $c$ directions, although $b$ itself was set to constant 1. From Fig. \ref{t0fig1}(d) it is clear that this force is always attractive in the entire range of parameters. Projecting the zero force boundary in Fig. \ref{t0fig1}(b), (c) onto the parameter space spanned by $(a,~c)$, one can clearly see where the force along $a,~b,~c$ directions are attractive or repulsive. One can also conclude that the force for all three pair of opposite faces cannot be made simultaneously repulsive \cite{Queiroz:2004wi}. In region I (or IV), the force along $a$ (or $c$) is repulsive while the other two directions are attractive. In region II, the force along both $a,~c$ are repulsive and that along $b$ direction is attractive. While in region III, the force along all directions are attractive.

\bc
\ec

The Eqs. \eqref{t0ce} and \eqref{t0cf} can also be used to obtain the limits in waveguide case (A2) and parallel plates case (A3). For the Casimir energy density and force density along $c$ direction per unit length of the waveguide, we obtain
\bea
F_{\rm w0}(b,c)&=&\lim_{a\rightarrow\infty}\frac{F_0(a,b,c)}{a}=-\biggl[ \frac{7\zeta(4)c}{32\pi^2 b^3}
+\frac{1}{b^{3/2}c^{1/2}}M_{\frac{3}{2}}\biggl(\frac{c}{b}\biggr) \biggr],
\label{t0wgce}\\
f_{\rm w0b}(b,c)&=&-\frac{\partial}{\partial b}F_{\rm w0}(b,c)
=-\frac{21\zeta(4)c}{32\pi^2 b^4}+\frac{2\pi c^{1/2}}{b^{7/2}}N_{\frac{1}{2}}\biggl(\frac{c}{b}\biggr),
\label{t0wgcf}
\eea
where Eq. \eqref{DK} has been used and
\be
N_{\frac{1}{2}}(x)=\sum_{k=0}\sum_{n=1}\frac {(-1)^{n+1}}{n^{1/2}}\Bigl(k+\frac{1}{2}\Bigr)^{5/2} K_{\frac{1}{2}}\Bigl(2\pi (k+\frac{1}{2})nx\Bigr). \label{N1/2}
\ee
The force density along $c$ direction takes the same form as Eq. \eqref{t0wgcf}  with $b$ and $c$ exchanged. In particular, for a waveguide with square cross-section, the Casimir energy and force along the two edges are
\bea
F_{\rm w0}(b,b)=-\frac{1}{b^2}\Bigl(\frac{7\pi^2}{2880}+0.0142\Bigr)=-0.0382\frac{1}{b^2},\label{Fw0bb}\\
f_{\rm w0b}(b,b)=\frac{1}{b^3}\Bigl(-\frac{7\pi^2}{960}+0.0338\Bigr)=-0.0382\frac{1}{b^3}\label{fw0bb}.
\eea
For the Casimir energy and force densities per unit area of the parallel plate, we have
\bea
F_{\rm p0}(b)&=&\lim_{c\rightarrow\infty}\frac{F_{\rm w0}(b,c)}{c}=-\frac{7\pi^2}{2880 b^3},
\label{t0ppce}\\
f_{\rm p0b}(b)&=&-\frac{\partial}{\partial b}F_{\rm p0}(b)=-\frac{7\pi^2}{960b^4}.
\label{t0ppcf}
\eea

\begin{figure}
(a)\hspace{7cm} (b)\\
\vspace{5mm}
\includegraphics[scale=0.37]{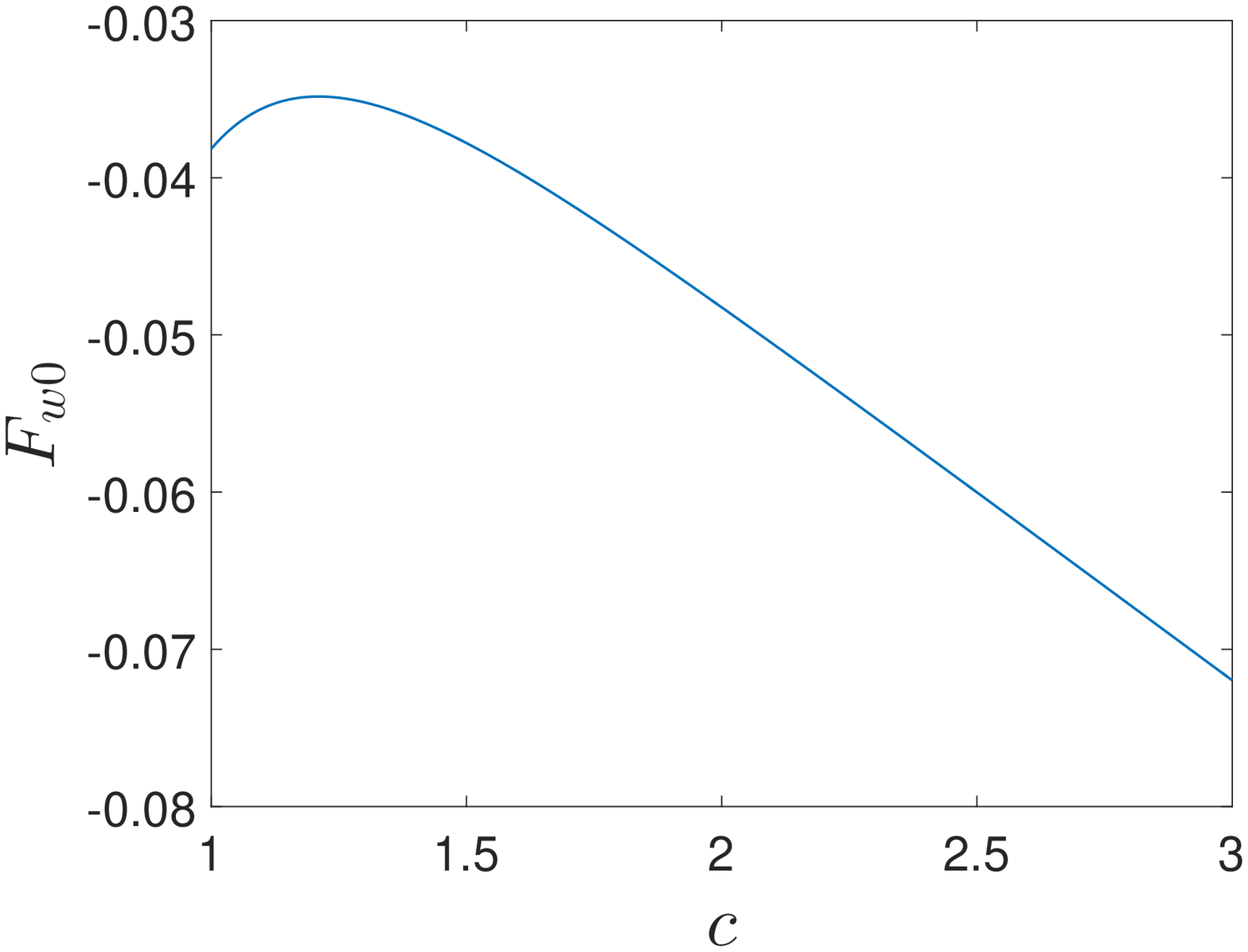}\hspace{4mm}
\includegraphics[scale=0.4]{fig3f-c.eps}
\caption{(a): Casimir energy density per unit length for a fermionic field in a waveguide as the aspect ratio changes; (b): Corresponding Casimir force density along $b$ and $c$  directions. \label{t0fig2}}
\end{figure}

To study in the (A2) case the effect of the aspect ratio of the  waveguide cross-section, we plotted in Fig. \ref{t0fig2} the Casimir energy \eqref{t0wgce} and force \eqref{t0wgcf} along $b$ and $c$ directions by varying the only variable $c$ from 1 to 3 (note that $b=1$ and $a\to\infty$ already). It is seen from Fig. \ref{t0fig2}(b) that as $c$ increases, the force in the $b$ direction $f_{\rm w0b}$ is always attractive. However, in Fig. \ref{t0fig2}(a) there exists a maximal point of the Casimir energy when the ratio $r_{\rm 0cr}=c/b\simeq 1.21$ which corresponds to a turning point of the force along $c$ direction ($f_{\rm w0c}$ in the Fig. \ref{t0fig2}(b)): when $c$ is below this value, $f_{\rm w0c}$ was attractive and above it,  $f_{\rm w0c}$ becomes repulsive. It is notable that Ref. \cite{Queiroz:2004wi} used the Bogoliubov transformation method and found a similar transformation but with different critical
aspect ratio $r_{\rm 0cr}$.

\bc
\ec

For the parallel plate case (A3), the results, Eqs. \eqref{t0ppce} and \eqref{t0ppcf}, are particularly simple. It is seen that the Casimir energy is always negative and monotonically increasing, while the force is always attractive, as anticipated from previous studies. These results are agree exactly with the Ref.~\cite{Fosco:2008vn, Gundersen:1987wz, Mostepanenko:1988bs, Milonni:1994xx, Queiroz:2004wi, Johnson, Santos:2002jp}.

\subsection{High temperature limit: $1/T\to0$ \label{caseb}}

This is another case for which the effective parameter space is further reduced and therefore easier to analyze. Similar to the low temperature limit in Sec. \ref{casea}, the Casimir energy in this limit should also depend on one length scale, e.g., $b$, and the ratios of other edges to $b$. Without losing generality therefore we also set $b=1$.

In high temperature however, as will be shown in Appendix~\ref{Appendix2} both the Casimir energy and Casimir force approach zero
\bea
&&\lim_{T\to\infty}F_{\rm{C}} =0, \label{lim Fc}\\
&& \lim_{T\to\infty}f_a =0. \label{lim fa}
\eea
These are in alignment with the effect of high temperature, whose thermal fluctuation will suppress the quantum fluctuation that is responsible for a finite Casimir energy and force.

\subsection{Finite temperature case}
In this case, we will use $1/T$ as the scale against which all edge sizes will be compared. For the purpose of studying Casimir energy and force,  without losing any generality we can simply set $T=1$ while allowing $b,~c,~a$ to vary freely in the reduced parameter space $(b\leq c\leq a)$. It is also clear that we do not need to study the case that all three edges are much larger than one, which is equivalent to high temperature case (\ref{casea}), or the case that all three edges are much smaller than one, which is equivalent to the low temperature case (\ref{caseb}). Taking all these into account, there are only 4 subcases that we need to study here: (C1) two edges $a$ and $c$ are much larger than $1/T$ while $b$ is comparable or smaller than $1/T$; (C2) one edge $a$ is much larger than $1/T$ while $c$ and $b$ are comparable or much smaller than $1/T$; (C3) $a$ and $c$ are comparable to $1/T$ while  $b$ is comparable or much smaller and finally (C4) $a$ is comparable to $1/T$ while $c$ and $b$ are much smaller.

\subsubsection{Case C1}

\begin{figure}[htp]
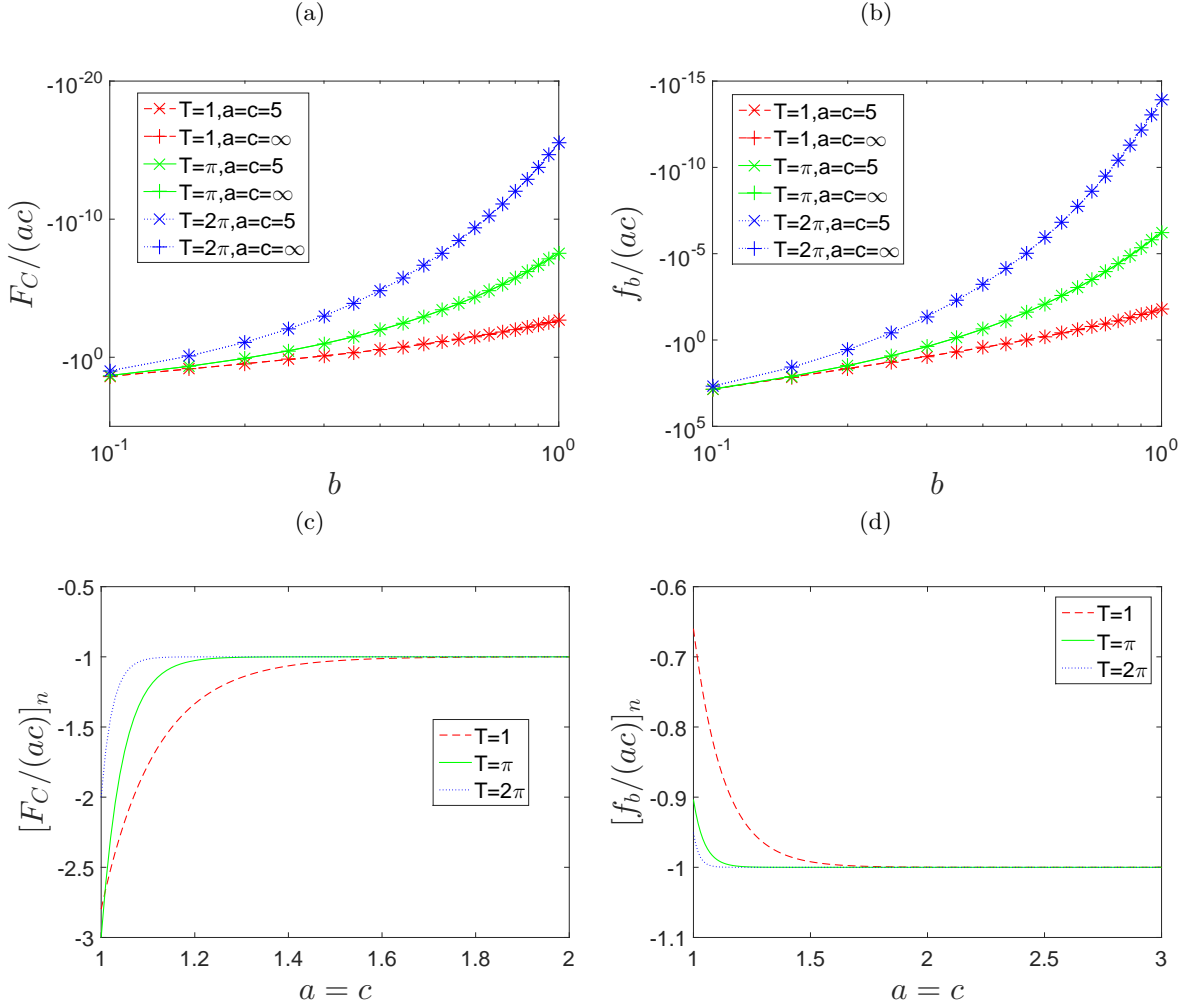

(a)\hspace{7cm} (b)\\
\vspace{5mm}
\includegraphics[scale=0.4]{fig4-E.eps}\hspace{4mm}
\includegraphics[scale=0.4]{mfig4-F.eps}\\
(c)\hspace{7cm} (d)\\
\vspace{5mm}
\includegraphics[scale=0.4]{fig4EaNm.eps}\hspace{4mm}
\includegraphics[scale=0.4]{mfig4fbaNm.eps}
\caption{(a): The Casimir energy density per unit area; (b) Casimir force density per unit area; (c) the edge size dependence of the normalized energy density; (d) the edge size dependence of the normalized force density.
\label{figc1}}
\end{figure}

Case (C1) is equivalent, in the limit that $a$ and $c$ are large, to a parallel plate geometry at finite temperature. The  Casimir energy and force densities per unit area in these limits are
\bea
F_{\rm p}(b,T)&=&\lim_{c\rightarrow\infty}\frac{F_{\rm w}(b,c,T)}{c}=-\frac{(2T)^{5/2} M_{\frac{3}{2}}(2bT)}{b^{1/2}}
\label{ppftce}\\
f_{\rm pb}(b,T)&=&-\frac{\partial}{\partial b}F_{\rm p}(b,T)=-\frac{2(2T)^{5/2}}{b^{1/2}}\biggl[\frac{1}{b}M_{\frac{3}{2}}(2bT)+2\pi T N_{\frac{1}{2}}(2Tb) \biggr],\label{ppftcf}
\eea
where $F_{\rm w}(b,c,T)$ is given in Eq. \eqref{wgftce} and $N_{\frac{1}{2}}(x)$ is given in Eq. \eqref{N1/2}.
We also compared these results with available literature and found that our Casimir energy density \eqref{ppftce} agrees with Eq. (3.17) of Ref. \cite{Gundersen:1987wz} (see Appendix \ref{Appendix3}) after subtracting the black body radiation term and changing its variables $\xi\to bT$. To see more clearly the dependence of the energy and force on the plate distance and temperature, we plot in Fig. \ref{figc1} in logarithmic scale the Casimir energy and force per unit area for two choices of edges $a=c=5$ and $a=c=\infty$ and three choices of temperature $T=1,~\pi$ and $2\pi$.

\bc
\ec

From Fig. \ref{figc1}(a), (b), one sees that both the Casimir energy and force density are always negative and increases to asymptotically zero as $b$ increases, which is expected because the plates distance increases. Note that in order to separate curves in the plots, a logarithmic scale was used in the $y$ axis. In the linear scale, both the Casimir energy and force densities are almost a constant zero as $b$ approaches 1. For the effect of temperature, it is seen that the higher the temperature, the faster the Casimir energy and force densities approach zero as $b$ increases. This is in agreement with the general effect of temperature increase, which always competes with that of the quantum fluctuations responsible for Casimir energy/force and therefore suppresses them. What is remarkable here is the effect of edge sizes to the Casimir energy and force density. As can be seen from plots Fig. \ref{figc1}(a), (b), the densities of both the Casimir energy and force completely coincides for $a=c=5$ and $a=c=\infty$. Indeed, in Fig. \ref{figc1}(c) and (d) we show how the energy and force densities normalized by their asymptotic magnitudes depend on the edge sizes. It is seen from the flat tails of these plots that a box with height  1 and square top and bottom faces with  edge larger than 2 has the same Casimir energy and force densities as a pair of infinitely large parallel plates with same plate distance. Moreover, the high the temperature the smaller the $a$ and $c$ need to be to resemble the asymptotic values of the Casimir force and densities. This is not surprising given that the higher temperature tends to demolish both the Casimir energy and force, as shown in the high temperature limits in subsection \ref{caseb}. This temperature effect to the Casimir energy and force densities is also observed here, although not shown in the normalized plot.

\subsubsection{Case C2}

\begin{figure}[htp]
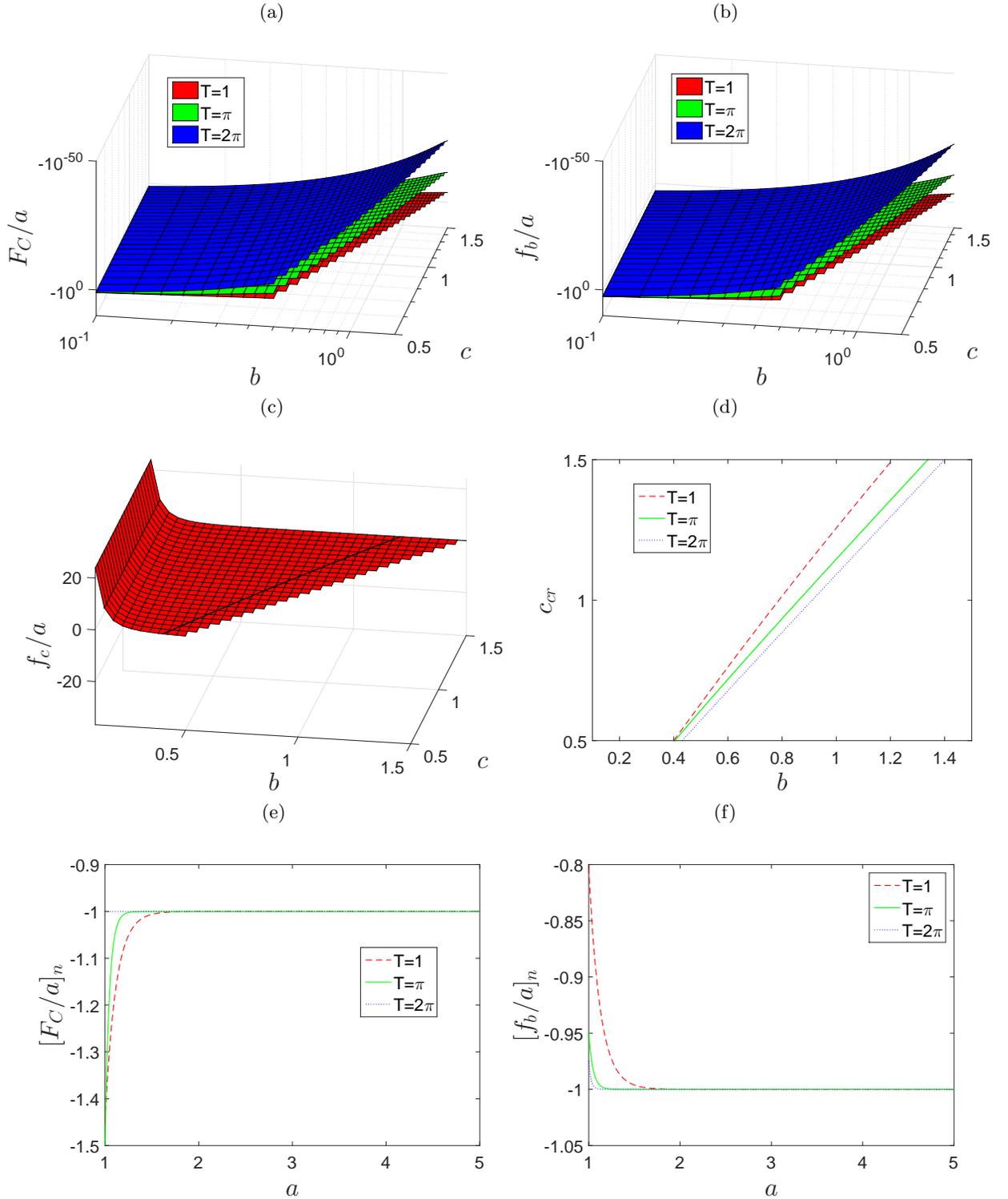

(a)\hspace{7cm} (b)\\
\vspace{5mm}
\includegraphics[scale=0.4]{fig5allTE.eps}\hspace{4mm}
\includegraphics[scale=0.4]{mfig5allTFb.eps}\\
(c)\hspace{7cm} (d) \\
\vspace{5mm}
\includegraphics[scale=0.4]{mfig5fcb.eps}\hspace{4mm}
\includegraphics[scale=0.4]{fig5contour2c.eps}\\
(e)\hspace{7cm} (f)\\
\vspace{5mm}
\includegraphics[scale=0.4]{fig5EaNm.eps}\hspace{4mm}
\includegraphics[scale=0.4]{mfig5fbaNm.eps}
\caption{(a): Casimir energy density per unit length at temperatures $T=1,~\pi,~2\pi$; (b): Casimir force density along $b$ direction at temperatures $T=1,~\pi,~2\pi$; (c): Casimir force density along $c$ direction at temperatures $T=1$; (d): the critical $c_{\rm{cr}}$ at which the force along $c$ direction flips sign at temperature changes; (e): the long-side edge dependence of the normalized energy and; (f): the long-side edge dependence of the normalized force. We set $a=5$. \label{figc2}}
\end{figure}

Case (C2) correspond to a waveguide geometry at high temperature compared to the waveguide's longest edge in the limit that $a$ is large. The Casimir energy and force density along $b$ direction per unit length in this limit are given by
\bea
F_{\rm w}(b,c,T)&=&\lim_{a\rightarrow\infty}\frac{F_{\rm{ C}}}{a}=-4T\biggl[ M_1\biggl(c,b,\frac{1}{2T}\biggr)
+c\biggl(\frac{2T^3}{b}\biggr)^{1/2}M_{\frac{3}{2}}(2bT) \biggr],
\label{wgftce}\\
f_{\rm wb}(b,c,T)&=&-\frac{\partial}{\partial b}F_{\rm w}(b,c,T)=8Tc\biggl\{\frac{\pi}{b^3}N_0\biggl(c,b,\frac{1}{2T}\biggr)
-(2T^3)^{\frac{1}{2}}\Bigl[ \frac{1}{b^{3/2}}M_{\frac{3}{2}}(2Tb)
+\frac{2\pi T}{b^{1/2}}N_{\frac{1}{2}}(2Tb) \Bigr]\biggr\}.\label{fwb} \label{wgftcf}
\eea
Eq. \eqref{wgftce} is derived in Appendix \ref{Appendix3}. Note that the force along $c$ direction takes the same form as Eq. \eqref{wgftcf} but with $b$ and $c$ exchanged. To see clearly the edge size and temperature dependence of these quantities, we plot them in Fig. \ref{figc2}  for some $c$ from smaller than 1 to comparable to 1, and $b$ from $b\ll 1$ to $c$ while fixing $a$ at 5. The increase of $b$ from $b\ll 1$ to $c$ is equivalent to the change the waveguide cross-section from a narrow rectangular to a square.

\bc
\ec

Fig. \ref{figc2}(a) shows that all Casimir energy densities are negative for all temperature and edge sizes. It  increases monotonically as $b$ increases in all range of $b\leq c$ and therefore the force along the $b$ direction is always attractive, as show in the  force plot, Fig. \ref{figc2}(b). As $c$ increases while keeping $b$ fixed however, a careful inspection shows that when $b$ is small, $b\leq c_{\rm min}\approx 0.5$, the Casimir energy monotonically decreases. This leads to a repulsive force along the $c$ direction. However,  for larger fixed $b$,  there exists a small interval of $c\in[b,~c_{\rm{cr}}]$ in which the Casimir energy increases as $c$ increases and after passing the critical $c_{\rm{cr}}$ the Casimir energy decreases again. This feature cannot be seen very clearly in Fig. \ref{figc2}(a) because of the finite element limit in it, but it is clearly shown in the $c$ direction force density plotted in Fig. \ref{figc2}(c). This means that by changing length of one side of the waveguide cross-section, the nature of the Casimir force can be changed.
As $b$ increases, the critical $c_{\rm{cr}}$ forms a curve in this plot. Therefore this curve is a  boundary in the parameter space spanned by $b$ and $c$, separating the attractive (right side of the line) and repulsive (left side of the line) forces along the $c$ direction. We also studied how this critical boundary depends on the temperature in Fig. \ref{figc2}(d). It is seen that the higher the temperature, the smaller the $c_{\rm{cr}}$ is required for the force to flip sign.

Finally, as for the long-edge size effect, similar to the previous case of parallel plate, we found that both the Casimir energy and force densities practically gain their asymptotic value when $a$ is as small as $2$ (See Fig. \ref{figc2}(e), (f) which show the normalized energy and force respectively). Besides, the value of $a$ at which the energy and force reach their asymptotic value also decreases as temperature increases, which is the same feature as in the C1 case and understandable given high temperature suppresses the quantum fluctuation responsible for the Casimir energy/force.

\subsubsection{Case C3}

\begin{figure}[htp]
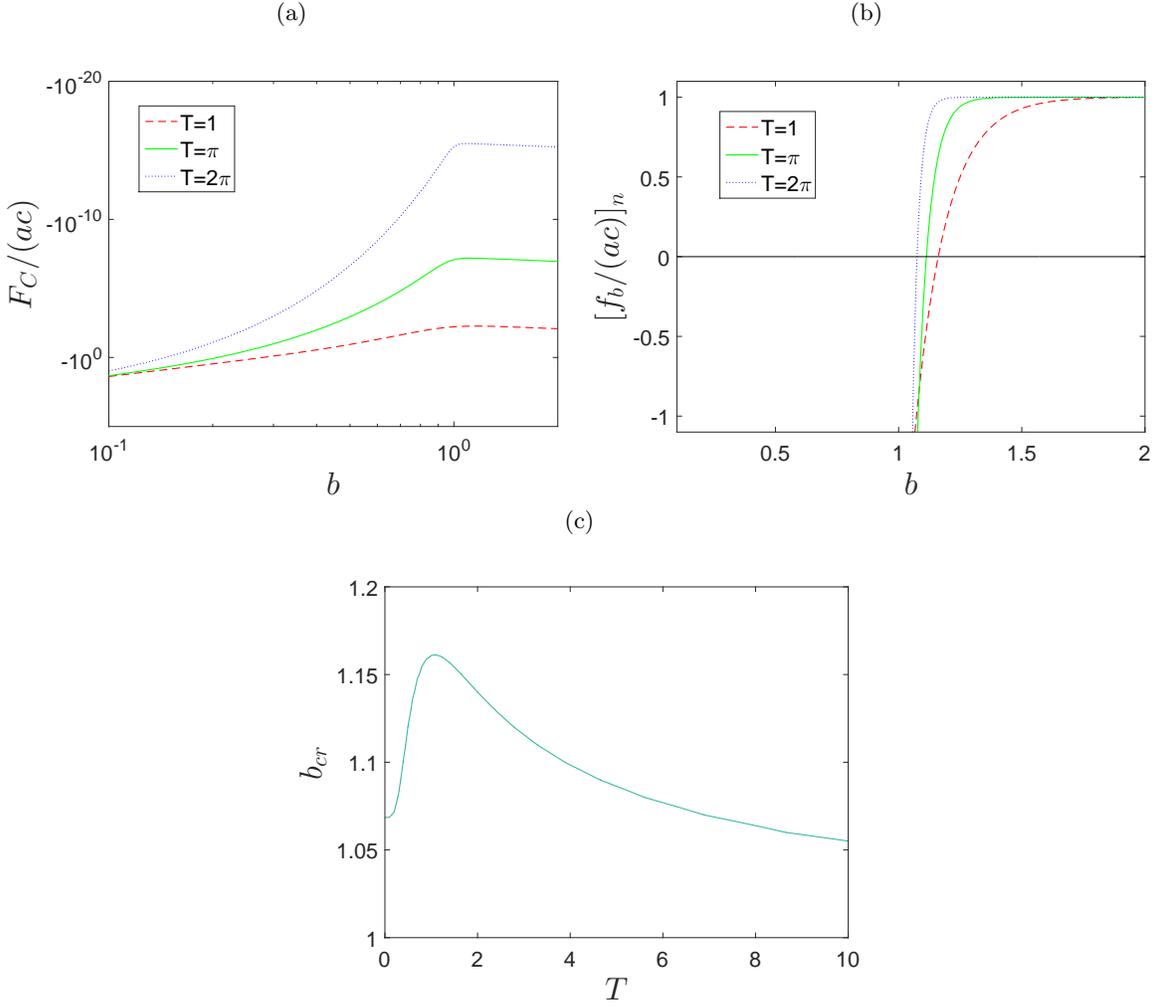

(a) \hspace{7cm} (b)\\
\vspace{5mm}
\includegraphics[scale=0.4]{mfig6E.eps}\hspace{4mm}
\includegraphics[scale=0.4]{mfig6fb6.eps}\\
(c)\\
\vspace{5mm}
\includegraphics[scale=0.4]{fig6Tb.eps}
\caption{(a): Casimir energy density and; (b): normalized Casimir force densities using Eq. \eqref{ppftce} and \eqref{ppftcf} for $a=c=1$ and $b$ from 0.1 to 2; (c): Temperature dependence of the critical length.  \label{figc3}}
\end{figure}

For case (C3), when $b$ is much smaller than $a$ and $c$, this is also equivalent to a parallel plate geometry, although the temperature here is kept low so that its inverse is comparable to the  plate edge sizes.  In limits that $a,~c$ are large, the Casimir energy and force have been given by Eqs. \eqref{ppftce} and \eqref{ppftcf}.
As $b$ increase, then the geometry becomes a typical rectangular box with all three edges comparable. We plotted in Fig. \ref{figc3} the Casimir energy and force density per unit area for a few temperatures while keeping $a=c=1$ and let $b$ varies from 0.1 to 2.

\bc
\ec

It is found from Fig. \ref{figc3}(b) that similar to case (C1), for all temperature as long as $b$ was smaller than $a$ and $c$, the Casimir force along $b$ direction will always be attractive. While as $b$ approaches $a$ and $c$ from below, the attractive force becomes weaker and approaches zero. After $b$ passing $a=c=1$, the geometry approaches a waveguide, which becomes similar to case (C2).  It is also found that the force along $b$ direction, which is now the longer direction of the waveguide, can also change from attractive to repulsive after passing a critical $b_{\rm{cr}}$. We also plotted the temperature dependence of this critical edge size in Fig. \ref{figc3}(c) and found that when temperature is higher than about $T=1.1$, then similar to the critical $b_{\rm{cr}}$ in case (C2), the critical $b_{\rm{ cr}}$ also decreases as the temperature increases. While for temperature below $T=1.1$, the critical $b_{\rm{cr}}$ increases with the increase of temperature.

\subsubsection{Case C4}

\begin{figure}
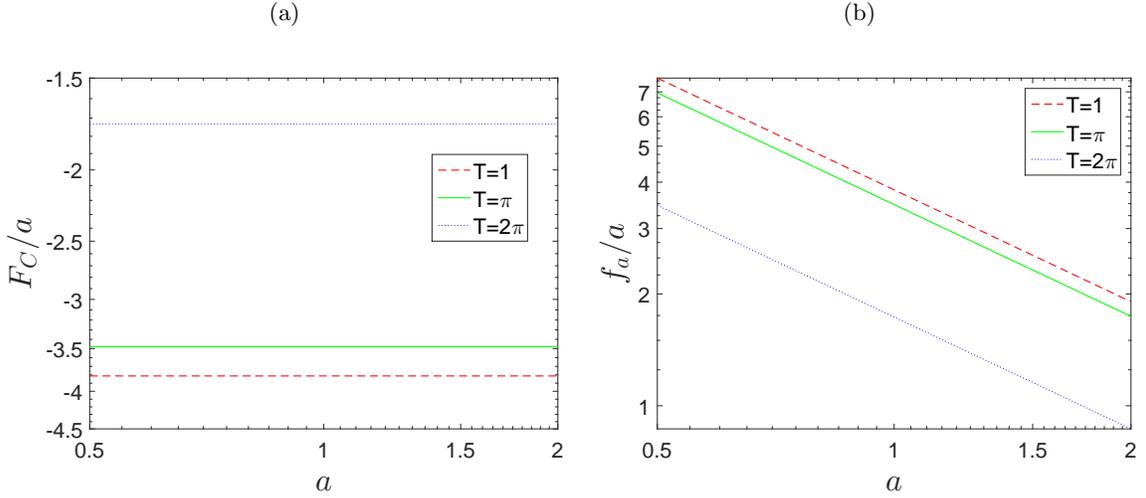

(a) \hspace{7cm} (b)\\
\vspace{5mm}
\includegraphics[scale=0.4]{fig7E.eps}\hspace{4mm}
\includegraphics[scale=0.4]{fig7fa.eps}\\
\caption{(a) Casimir energy density per unit length along $a$ direction; (b) the corresponding force density. We set $T=1,~\pi,~2\pi$ and $b=c=0.1$. \label{figc4}}
\end{figure}

For case (C4), the geometry is similar to a waveguide case but the temperature is comparable to $1/a$ which is in contrast to case (C2), and much smaller than $1/b$ or $1/c$. This also requires us to not take the $a\to\infty$ limit. We plotted in Fig. \ref{figc4} the Casimir energy and force density per unit length along the $a$ direction by setting $T=1,~\pi,~2\pi$ and $b=c=0.1$ and varying $a$ from $1/2$ to $2$. This force density is along the longitudinal direction and analogous to the spring factor in Hooke's law.  Therefore, it describes how the force factor changes with respect to the waveguide length.

It is seen from the plots that under such large length to width ratio, the Casimir energy exhibit an expected behavior that its density is a constant meaning the total Casimir energy is proportional to the length. This is similar to a pair of large parallel plate in that both are proportional to the large dimension of the geometry. The foundation of this proportionality of course is that the shortest edge(s) of the rectangular box determines the density of Casimir energy, be it per unit area or per unit length. This constant energy density then means that the force density along $a$ direction, i.e., the force factor, decreases as $1/a^1$. This is seen in Fig. \ref{figc4}(b) and also easily understand from the Hooke's law.

\bc
\ec

\begin{figure}
\includegraphics[scale=0.4]{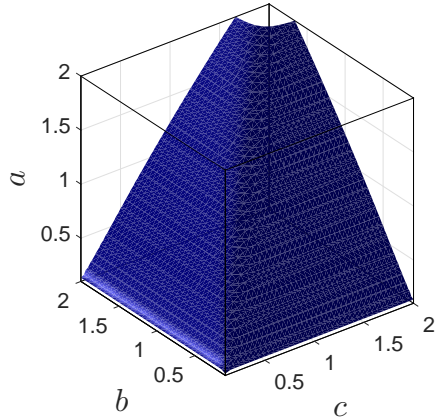}
\caption{Transition boundary for the force $f_a$ from attractive to repulsive at $T=1$. \label{figzf}}
\end{figure}

Summarizing cases (C1) to (C4) and to get a better understanding of the transition of the Casimir force from attractive or repulsive, we combine the analysis done in the above four cases, and plot in Fig. \ref{figzf} the transition surface in the parameter space spanned by all three edge size $a,~b$ and $c$ from 0.1 to 2 for temperatures $T=1$ for the force along $a$ direction. It is seen that for a fixed and small $a$, there exists an L-shaped boundary composed mainly by two straight lines at small $b$ and small $c$ respectively. In one side of the boundary where $b$ and $c$ are simultaneously large, the force is attractive; while on the other side of the boundary, the force is repulsive. As $a$ increases, this L-shaped boundary also shrinks towards the larger $b$ and $c$ direction and eventually approaches $b,c\gtrsim 1.7$ when $a$ reaches 2.
\bc
\ec

\section{Discussions \label{secdis}}

The thermal Casimir energy and force for massless fermionic field confined in rectangular box are calculated in this paper. The analytic expressions are given in Eqs. \eqref{FC} and \eqref{fa}. Their various limiting values agree with previously known results in simpler geometries. Using these results, low and high temperature limits and effects of finite temperature and box edge sides on the Casimir energy and force were studied. Generally, it is found that at zero temperature, there exist two boundaries (see Fig. \ref{t0fig1}) dividing the effective two-dimensional parameter space into four regions. In one of the regions all forces along three edges are attractive while in two other regions the force along the longest edge becomes repulsive and in the last region forces along two longest edge becomes positive. For the finite temperature case, the parameter space is divided into four subcases. For a box with geometry similar to parallel plate and high temperature, the force between the plate is always attractive and becomes weaker as the plate distance or temperature increases. For the waveguide geometry at high temperature, depending on the aspect ratio of the waveguide cross-section, the forces along the wider side of the cross-section can transform from attractive to repulsive. The transition value of the longer cross-section edge decreases as temperature increases. For geometry of parallel plate with low temperature or geometry of box with three comparable sizes, there also exist a critical value for the longest edge length beyond which the force along this direction changes from attractive to repulsive. This critical value changes with temperature non-monotonically. For the waveguide geometry at low temperature, as the length of the waveguide increases, the Casimir energy density per length is kept at a constant and the force density per length along the longitudinal direction decreases as length inverse. It is found that at general temperature, the parameter space of three edges can always be split into by a surface into two regions according to the nature of the Casimir force along any particular direction. In the high temperature limit, it is found that both the Casimir energy and force approaches zero.

As for the extension of the current work, two possible choices might be attempted. The first is to consider other boundary conditions. Although bag model boundary conditions make the solution of the frequency modes simple, there do exist other boundary conditions \cite{Zhai:2011zza,Elizalde:2011cy}, which might be more suitable for the system one wants to study. The second is to consider a massive fermionic field, which will introduce another energy scale against which the effect of temperature and edge sizes can be compared. Moreover, fermions with nonzero mass is more realistic given that the Casimir effect experiments are always carried out in condensed matter systems, in which the fermionic excitation (quasi-particles) usually have a non-zero (although sometimes small) mass. For these two directions, we expect the later shall be easier because the former will alter the modes of the allowed quantum fluctuation and therefore affects computation in a more fundamental way.

A more dramatic turn of the future direction would be studying thermal Casimir effect of Dirac, Majorana or Weyl fermions in three-dimensional box. With the rise of these kinds of fermionic quasi-particles in condensed matter system in the past years, there has also been studies of fermionic Casimir effects of them \cite{fgraphene, fgraphene2, fgraphene3}. However, there is still a lacking of studies for more complex geometry such as a three-dimensional box using these fermions with arbitrary temperature. We are currently working in this direction.

Finally, in this work we extended the Schl\"omilch's formula to the cases of double series and triple series. These generalized formulas can be applied to the calculation of the thermal Casimir effect for scalar field confined in rectangular boxes \cite{Geyer:2008wb}.

\begin{acknowledgments}
The authors appreciate the help of Mr. Moran Qin in numerical verifications of some equations. This research is supported by the Chinese SRFDP 20130141120079, NNSF China 11504276 \& 11547310 and MST China 2014GB109004.
\end{acknowledgments}

\appendix
\section{Deriving the expression of $F_T$ \label{appd1}}
In this appendix, we show how Eq.~\eqref{FT} can be derived using Eq.~\eqref{three partition plus} with the help of an operator $\hat{S}$ defined below.

According to the definition of $F_T$ in Eq.~\eqref{infinite fermionic energy}, we have
\be
F_T=-4T\sum_{n,m,j=0}\ln\Bigl(1+e^{-\frac{\pi}{T}\sqrt{(\frac{2n+1}{2a})^2+(\frac{2m+1}{2b})^2+(\frac{2j+1}{2c})^2}}\Bigr).
\ee
The summations can be recast into the form
\be
F_T=4T\hat{S}U(a,b,c,T), \label{V-U}
\ee
where the operator $\hat{S}$ is defined by its action on any function $u(a,b,c)$ as
\bea
\hat{S} u(a,b,c)
&=&u(2a,2b,2c)+u(a,b,2c)+u(a,2b,c)+u(2a,b,c)\nonumber\\
&&-u(a,2b,2c)-u(2a,b,2c)-u(2a,2b,c)-u(a,b,c), \label{operation}
\eea
and $U(a,b,c,T)$ in Eq. \eqref{V-U} is defined as
\be
U(a,b,c,T)=\sum_{n,m,j=1}\ln\Bigl(1+e^{-\frac{\pi}{T}\sqrt{(\frac{n}{a})^2+(\frac{m}{b})^2+(\frac{j}{c})^2}}\Bigr).
\ee

The quantity $U(a,b,c,T)$ can be calculated by using Eq.~\eqref{three partition plus} and the result is
\bea
U(a,b,c,T)&=&U_1(a,b,c,T)-\biggl[\frac{7\zeta(4)abcT^3}{8\pi^2} +\frac{\zeta(4)ac}{16\pi^2 b^3 T}\biggr]
-\frac{\pi}{T}\biggl[\frac{a}{4\pi b^{3/2}c^{1/2}}Y_{\frac{3}{2}}\biggl(\frac{c}{b}\biggr)+\frac{1}{2\pi}V_1(a,b,c) \biggr]\nn\\
&&
+Z_3\biggl(a,b,c,\frac{1}{2T}\biggr)-Z_3\biggl(a,b,c,\frac{1}{T}\biggr)
-aV_1\biggl(c,b,\frac{1}{2T}\biggr)+aV_1\biggl(c,b,\frac{1}{T}\biggr)\nn\\
&&
-ac\Bigl(\frac{2T^3}{b}\Bigr)^{1/2}Y_{\frac{3}{2}}(2Tb)+\frac{ac}{2}\Bigl(\frac{T^3}{b}\Bigr)^{1/2}Y_{\frac{3}{2}}(Tb). \label{U combination}
\eea
Here
\bea
U_1(a,b,c,T)&=&\frac{3\zeta(3)abT^2}{16\pi}
+\frac{3\zeta(3)acT^2}{16\pi}
-\frac{\pi aT}{48}
+\frac{\zeta(3)a}{32\pi b^2 T}
+\frac{1}{2}\biggl[Z_2\biggl(\frac{1}{T},b,c\biggr)
-Z_2\biggl(\frac{1}{2T},b,c\biggr)\biggr]\nonumber\\
&&-\frac{aT Y_1(bT)}{2}
+aT Y_1(2bT), \label{U4}
\eea
$Y_{3/2}(x)$ and $V_1(x,y,z)$ were defined in Eq.~\eqref{Y3/2-V1}, and $Z_3(x,y,z)$ were defined in Eq. \eqref{Z3}.

The operator $\hat{S}$ is linear, and have the following properties when applied onto functions with one, two or three variables with special form
\bea
&&\hat{S}u(a)=\hat{S}u(b)=\hat{S}u(c)=\hat{S}u(a,b)=\hat{S}u(a,c)=\hat{S}u(b,c)=0,\nonumber\\
&&\hat{S}[a\cdot U(b,c)]=a[u(2b,2c)+u(b,c)-u(b,2c)-u(2b,c)],\nonumber\\
&&\hat{S}[ac\cdot u(b)]=ac[u(2b)-u(b)].\label{hatspro1}
\eea
Therefore when it is applied to each term in $U(a,b,c,T)$ in Eq. \eqref{U combination}, we have
\bea
&&\hat{S}U_1(a,b,c,T)=0, \label{hatU4}\\
&&
\hat{S}\biggl[-\frac{7\zeta(4)abc T^3}{8\pi^2}-\frac{\zeta(4)ac}{16\pi^2 b^3 T}\biggr]
=-\frac{7\zeta(4)abc T^3}{8\pi^2}
+\frac{7\zeta(4)ac}{128\pi^2 b^3 T}, \label{hatU1}\\
&&\hat{S}\biggl[\frac{aY_{3/2}(c/b)}{b^{3/2}c^{1/2}} \biggr]
=-\frac{a}{b^{3/2}c^{1/2}}M_{\frac{3}{2}}\biggl(\frac{c}{b}\biggr), \label{hat u22}\\
&&\hat{S}V_1(a,b,c)=-M_1(a,b,c), \label{hat m1}\\
&&
\hat{S}\biggl[ Z_3\biggl(a,b,c,\frac{1}{2T}\biggr)-Z_3\biggl(a,b,c,\frac{1}{T}\biggr)\biggr]
=-W_3\biggl(a,b,c,\frac{1}{2T}\biggr), \label{hat u31}\\
&&
\hat{S}\biggl[-aV_1\biggl(c,b,\frac{1}{2T}\biggr)
+aV_1\biggl(c,b,\frac{1}{T}\biggr) \biggr]
=-aM_1\biggl(c,b,\frac{1}{2T}\biggr), \label{hat-u32}\\
&&
\hat{S}\biggl[-ac\biggl(\frac{2T^3}{b}\biggr)^{1/2}Y_{\frac{3}{2}}(2Tb)
+\frac{ac}{2}\biggl(\frac{T^3}{b}\biggr)^{1/2}Y_{\frac{3}{2}}(Tb) \biggr]
=-ac\biggl(\frac{2T^3}{b}\biggr)^{1/2}M_{\frac{3}{2}}(2bT), \label{hat-u33}
\eea
where $M_{\frac{3}{2}}(x)$, $M_1(x,y,z)$ and $W_3(x,y,z,t)$ were defined in Eqs. \eqref{M3/2}, \eqref{M1} and \eqref{W3} respectively.

Finally, substituting them back into Eq. \eqref{U combination} and \eqref{V-U} yield Eq. \eqref{FT} in Sec.~\ref{Fermion field}.

\section{High temperatures limits\label{Appendix2}}
In this section we first prove Eq.~\eqref{E physic_ht_old} in Sec.~\ref{EM field}. In order to do so, we only need to show that the $F_2(a,b,c,T)$ term in Eq. \eqref{F2(abl-alpha)} approaches zero.
Let us prove term by term that this will be zero in the high temperature limit. When $T$ is high enough, according to definition~\eqref{Z3}, we have
\be
0<-T Z_3\biggl(a,b,c,\frac{1}{2T}\biggr)
<T\sum_{k,m,j}e^{-\pi a\sqrt{\frac{m^2}{b^2}+\frac{j^2}{c^2}+4T^2k^2}}
<T\sum_{k,m,j}e^{-\pi a\frac{1}{\pi}(\frac{m}{b}+\frac{j}{c}+2Tk)}
 \label{inequlity 1}
\ee
Since the right side of Eq.~\eqref{inequlity 1} approaches zero at high temperatures, we have
\be
\lim_{T\rightarrow\infty}TZ_3\biggl(a,b,c,\frac{1}{2T}\biggr)=0^-. \label{lim 1}
\ee
Similarly, using definition \eqref{Z1-Z2} we can prove
\be
\lim_{T\rightarrow\infty}TZ_2\biggl(x,y,\frac{1}{2T}\biggr)=0^-,
\lim_{T\rightarrow\infty}TZ_1\biggl(2Tx\biggr)=0^-. \label{lim 2}
\ee

The asymptotic expression of the Bessel functions of an imaginary argument at limit $x\rightarrow\infty$ is~\cite{Masirevic}
\be
K_\nu(x)=e^{-x}\sqrt{\frac{\pi}{2x}}\bigl[1+O(x^{-1})\bigr],\quad x\rightarrow\infty. \label{asymptotic K}
\ee
According to Eqs.~\eqref{Y3/2-V1} and~\eqref{asymptotic K} then, when $T\rightarrow\infty$,
\bea
&&TV_1\biggl(c,b,\frac{1}{2T}\biggr)
\sim T\sum_{k,m,n}\frac{\sqrt{4T^2k^2+\frac{m^2}{b^2}}}{n}\sqrt{1/\Bigl(4nc\sqrt{4T^2k^2+\frac{m^2}{b^2}}\Bigr)}e^{-2\pi cn\sqrt{4T^2k^2+\frac{m^2}{b^2}}}\nonumber\\
&&< T\sum_{k,m,n}\Bigl(2Tk+\frac{m}{b}\Bigr)e^{-2\pi cn\cdot \frac{1}{2\pi}(2Tk+\frac{m}{b})}
=T\sum_{k,m}\frac{2Tk+\frac{m}{b}}{e^{c(2Tk+\frac{m}{b})}-1}
<T\sum_{k,m}\frac{2Tk+\frac{m}{b}}{e^{c(2Tk+\frac{m}{b})}-\frac{1}{2}e^{c(2Tk+\frac{m}{b})}}. \label{inequlity 2}
\eea
It is clear then
\be
\lim_{T\rightarrow\infty}TV_1\biggl(c,b,\frac{1}{2T}\biggr)=0^+. \label{lim M1}
\ee

According to Eqs.~\eqref{X0-Y1} and~\eqref{asymptotic K}, when $T\rightarrow\infty$,
\be
T^2Y_1(2cT) \sim \sqrt{\frac{T^3}{8c}}\sum_{k,n}\sqrt{\frac{k}{n^3}}e^{-4\pi Tckn}
< \sqrt{\frac{T^3}{8c}}\sum_{k,n}ke^{-4\pi Tckn}
=\sqrt{\frac{T^3}{8c}}\sum_n \frac{e^{4\pi Tcn}}{(e^{4\pi Tcn}-1)^2}. \label{inequlity 4}
\ee
Thus, it is also clear that the exponential term in the denominator will win over and therefore
\be
\lim_{T\rightarrow\infty}T^2Y_1(2cT)= 0^+. \label{lim M2}
\ee

According to Eq.\eqref{Y3/2-V1} and formula~\cite{Gradshteyn}
\bea
K_{i+1/2}(z)=\sqrt{\frac{\pi}{2z}}e^{-z}\sum_k^i\frac{(i+k)!}{k!(i-k)!(2z)^k}, \label{K3/2}
\eea
when integer $i$ equals 3, one can derive
\bea
T^{5/2}Y_{\frac{3}{2}}(2bT)
=\frac{T}{2^{7/2}\pi b^{3/2}} \biggl[
4\pi Tb\sum_n\frac{e^{4\pi Tbn}}{n^2(e^{4\pi Tbn}-1)^2}+ \sum_n\frac{1}{n^3(e^{4\pi Tbn}-1)} \biggr]. \label{equlity 6}
\eea
Similar to the situation in Eq. \eqref{inequlity 4}, hence
\bea
\lim_{T\rightarrow\infty}T^{5/2}Y_{\frac{3}{2}}(2bT)= 0^+. \label{lim M3}
\eea

Finally, combining Eqs.~\eqref{lim 1},~\eqref{lim 2},~\eqref{lim M1},~\eqref{lim M2} and~\eqref{lim M3}, it follows then
\bea
\lim_{T\rightarrow\infty}F_2(a,b,c,T)= 0. \label{lim F2}
\eea
From this, Eq. \eqref{E physic_ht_old} can be immediately obtained.

Now let us prove the Eq. \eqref{lim Fc} and \eqref{lim fa} in Sec.~\ref{Fermion field}.
For the Casimir energy, according to Eqs.~\eqref{FC}, ~\eqref{A3}, ~\eqref{hat u22},~\eqref{hat m1}, \eqref{hat u31}, and taking into account definition~\eqref{operation}, Eqs.~\eqref{lim 1},~\eqref{lim M1},~\eqref{lim M3}, we can obtain very simply
\bea
\lim_{T\rightarrow\infty}F_{\rm{C}}&=&\lim_{T\rightarrow\infty}T A_3(a,b,c,T)
=\lim_{T\rightarrow\infty}T\hat{S}\biggl\{
Z_3\biggl(a,b,c,\frac{1}{2T}\biggr)-Z_3\biggl(a,b,c,\frac{1}{T}\biggr)
-aV_1\biggl(c,b,\frac{1}{2T}\biggr)\nonumber\\
&&+aV_1\biggl(c,b,\frac{1}{T}\biggr)
-ac\biggl(\frac{2T^3}{b}\biggl)^{1/2}Y_{\frac{3}{2}}(2Tb)+\frac{ac}{2}\biggl(\frac{T^3}{b}\biggr)^{1/2}Y_{\frac{3}{2}}(Tb) \biggr\}=0, \label{lim A3}
\eea
which is Eq. \eqref{lim Fc}.

For the Casimir force, denoting the first term of Eq.~\eqref{fa} by $f_1$,
it is clear that in the first term the exponential term in the denominator dominates the numerators
\be
\lim_{T\rightarrow\infty}f_1=0^-.\label{inequality of f1}
\ee
According to Eqs. \eqref{hat-u32} and \eqref{lim M1}, \eqref{hat-u33} and \eqref{lim M3}, and the definition of $\hat{S}$ in Eq. \eqref{operation}, one finds
\be
\lim_{T\rightarrow\infty}TM_1\biggl(c,b,\frac{1}{2T}\biggr)=\lim_{T\rightarrow\infty}T^{\frac{5}{2}}M_{\frac{3}{2}}(2Tb)=0.\label{lim f3}
\ee
Combination of Eq. \eqref{inequality of f1} and \eqref{lim f3} proves Eq.~\eqref{lim fa} in Sec.~\ref{Fermion field}.

\section{Waveguide and Parallel plate limits\label{Appendix3}}

In this Appendix we derive the formula for the Casimir energy per unit length in the case of waveguide and per unit area in the case of parallel plate, i.e., Eqs.~\eqref{wgftce} and~\eqref{ppftce}.

For the waveguide case, from definition Eqs.~\eqref{FC}, \eqref{A3} and \eqref{W3}, it is seen that in order to prove Eq. \eqref{wgftce}, we need to study the limits $\displaystyle \lim_{a\to\infty}W_3(a,b,c,x)/a$.
Similar to the argument from Eq.~\eqref{inequlity 1} to Eq.~\eqref{lim 1}, one can obtain
\be
\lim_{a\rightarrow\infty}\frac{1}{a}W_3\biggl(a,b,c,\frac{1}{2T}\biggr)=0. \label{limW3}
\ee
Using this, Eq. \eqref{wgftce} immediately follows.

In order to further prove Eq. \eqref{ppftce}, we need to study the limit of $\displaystyle \lim_{c\to\infty}M_1(c,x,y)/c$.
Similar to the argument from Eq.~\eqref{asymptotic K} to Eq.~\eqref{lim M1}, one can obtain
\be
\lim_{a\rightarrow\infty}\frac{1}{a}V_1(a,b,c)=0, \label{lim-u23a}
\ee
Then according to Eq. \eqref{hat m1} and definition of $\hat{S}$ in Eq. \eqref{operation}, Eq.~\eqref{lim-u23a} further implies
\bea
\lim_{a\rightarrow\infty}\frac{1}{a}M_1(a,b,c)=0. \label{limhatu23a}
\eea
Using this equation and Eq. \eqref{wgftce}, Eq. \eqref{ppftce} follows.

Lastly, let us show that after subtracting from Eq.~(3.17) of Ref.~\cite{Gundersen:1987wz} a free black body radiation energy term, the free Casimir energy will agree with our result \eqref{limhatu23a}.

We will do the proof backward. First letting $\xi=bT$ and defining
\be
g(\xi)=b^3 F_p(b,T),\label{definition-g}
\ee
then according to Eqs. \eqref{ppftce}, \eqref{M3/2} and \eqref{K3/2} this Casimir energy becomes
\bea
g(\xi)&=&-(2\xi)^{5/2}M_{\frac{3}{2}}(2\xi)
=\frac{\xi}{4\pi}\sum_n\frac{(-1)^n[\sinh(2\pi\xi n)+2\pi\xi n\cdot \cosh(2\pi\xi n)]}{n^3\cdot \sinh^2(2\pi\xi n)}\nonumber\\
&=&\frac{\xi}{4\pi}\biggl\{ \sum_n\frac{2[\sinh(4\pi\xi n)+4\pi\xi n\cdot \cosh(4\pi\xi n)]}{(2n)^3\sinh^2(4\pi\xi n)}-\sum_n\frac{\sinh(2\pi\xi n)+2\pi\xi n\cdot \cosh(2\pi\xi n)}{n^3\sinh^2(2\pi\xi n)} \biggr\}\nonumber\\
&=&-\frac{\xi^3}{16\pi}\frac{\partial}{\partial\xi}\frac{1}{\xi}\sum_n\frac{1}{n^3}\biggl( \frac{1}{\sinh(4\pi n\xi )}-\frac{4}{\sinh(2\pi n\xi)} \biggr).\label{g-xi}
\eea
Clearly, this is different from the dimensionless free energy $f(\xi)$ in Eq.~(3.17) of Ref.~\cite{Gundersen:1987wz}
by just the black body radiation term given in the square brackets below
\be
g(\xi)=f(\xi)-\left[-\frac{7\pi^2\xi^4}{180}\right]. \label{g-f}
\ee

\end{document}